\documentclass[review]{elsarticle}

\usepackage{lineno}
\modulolinenumbers[5]

\journal{arXiv}

\biboptions{authoryear}

\usepackage{color}

\usepackage{array}
\usepackage{enumitem}
\usepackage{amssymb}
\usepackage[colorlinks,urlcolor=blue]{hyperref}
\usepackage{amsmath}

{\left\lbrace\begin{array}{@{}l@{}}}%
{\end{array}\right.}

\begin{document}

\begin{frontmatter}

\title{A comprehensive framework for hard-magnetic beams: reduced-order theory, 3D simulations, and experiments}


\author[mymainaddress]{Dong Yan}
\ead{dong.yan@epfl.ch}

\author[mymainaddress]{Arefeh Abbasi}
\ead{arefeh.abbasi@epfl.ch}

\author[mymainaddress]{Pedro M. Reis\corref{mycorrespondingauthor}}
\cortext[mycorrespondingauthor]{Corresponding author}
\ead{pedro.reis@epfl.ch}

\address[mymainaddress]{Flexible Structures Laboratory, Institute of Mechanical Engineering\\
	\'{E}cole Polytechnique F\'{e}d\'{e}rale de Lausanne,
	1015 Lausanne, Switzerland\\
    }

\begin{abstract}
Thin beams made of magnetorheological elastomers embedded with hard magnetic particles (hard-MREs) are capable of large deflections under an applied magnetic field. We propose a comprehensive framework, comprising a beam model and 3D finite element modeling (FEM), to describe the behavior of hard-MRE beams under both uniform and constant gradient magnetic fields. First, based on the Helmholtz free energy of bulk (3D) hard-MREs, we perform dimensional reduction to derive a 1D description and obtain the equilibrium equation of the beam through variational methods. In parallel, we extend the existing 3D continuum theory for hard-MREs to the general case of non-uniform fields by incorporating the magnetic body force induced by the field gradient and implementing it in FEM. The beam model and FEM are first validated using experiments and then employed to predict the deflection of a cantilever beam in either a uniform or a constant gradient field. The corresponding parameters governing the magneto-elastic coupling are identified. Then, a set of comparative numerical studies for actuation in different configurations yields additional insight into the beam response. Our study builds on previous work on hard-MRE beams, while providing a more complete framework, both in terms of the methodologies used and the configurations considered, to serve as a valuable predictive toolbox for the rational design of beam-like hard-magnetic structures.
\end{abstract}


\end{frontmatter}


\section{Introduction}
\label{sec:Introduction}
Magnetorheological elastomers (MREs) are composite materials made of an elastomeric matrix embedded with magnetic particles; a composition that confers them the possibility of being simultaneously magnetically active and mechanically deformable. Subjected to an applied magnetic field, magnetic torques and/or forces are imposed on the particles and transmitted over the soft matrix, thereby inducing deformation of the composite. This coupling between mechanics and magnetism in MREs has been recently exploited to design functional devices for various applications~\citep{li_state---art_2014,gray_review_2014,hines_soft_2017,wu_multifunctional_2020}. In terms of the magnetic response of the embedded particles against applied fields, this class of composites can be classified as either soft- or hard-MREs, the characteristics of which will be discussed next, so as to better contextualize and help identity the niche of our study.

In soft-MREs, soft-ferromagnetic particles are prone to change their magnetization with an external field~\citep{Miyazaki_Ferromagnetism2012} and, due to their magnetic interactions, tend to form chains in a compliant elastomeric matrix aligned along the field direction~\citep{ginder_magnetostrictive_2002,danas_experiments_2012}. The displacements of the constituent micro-scale particles result in macro-scale deformations and in the change in the elastic properties of the composite~\citep{rigbi_response_1983,ginder_magnetorheological_1999,ginder_magnetostrictive_2002,danas_experiments_2012}. Leveraging this magneto-mechanical coupling effect, applications have been developed for tunable vibration absorbers~\citep{ginder_magnetorheological_2001}, microfluidic pumps and mixers~\citep{tang_versatile_2018,gray_review_2014}, and force sensors~\citep{li_development_2009}.

By contrast, in hard-MREs, hard-ferromagnetic particles, upon field saturation, possess a sufficiently high coercivity to resist demagnetization by external fields~\citep{Miyazaki_Ferromagnetism2012}. Consequently, the remnant magnetic moment of hard-MREs can be retained during actuation. In particular, flexible slender structures made out of hard-MREs are capable of significant shape changes, driven by the magnetic body torque induced by the interaction between the intrinsic magnetization of the material and an applied field~\citep{Lum_PNAS2016,Kim_Nature2018}. This fast, reversible, and remotely-controlled shape-shifting behavior has enabled novel functionalities in soft robotics~\citep{Hu_Nature2018,Gu_NatCommun2020,pancaldi_flow_2020}, biomedicine~\citep{Kim_SciRobot2019}, metamaterials~\citep{chen_reprogrammable_2021,montgomery_magneto-mechanical_2021}, and micromachines~\citep{alapan_reprogrammable_2020}. Moreover, the magnetization profile of a structure can be programmed by the local orientation of the magnetized particles to generate complex 3D shape transformations~\citep{Lum_PNAS2016,Kim_Nature2018,alapan_reprogrammable_2020}. At the microstructural level, many studies have focused on the particle-particle interaction in hard-MREs, investigating its influence on the magnetization and deformation of the composite~\citep{vaganov_effect_2018,vaganov_training_2020,schumann_reversible_2021,zhang_micromechanics_2020,garcia-gonzalez_microstructural_2021}. 

To rationally design magneto-active systems, there has been a recent research drive to devise predictive models for the mechanical behavior of soft- and hard-MREs, in response to an applied magnetic field. A review on the \textit{continuum-based} and \textit{micromechanically-based} modeling of soft-MREs can be found in~\cite{danas_experiments_2012} and~\cite{Zhao_JMPS2019}. The mechanics of soft-magnetic \textit{structural elements} has also been studied in the literature; \textit{e.g.}, beams~\citep{moon_magnetoelastic_1968,cebers_bending_2004,dreyfus_microscopic_2005,gerbal_refined_2015}, thin films~\citep{Psarra_JMPS2019}, and shells~\citep{Loukaides_IntJSmartNanoMater2014,Seffen_SmartMaterStruct2016}.

Since the center of the present study is on hard-magnetic beams, next, we will focus on recent advances in the continuously modeling of hard-MREs. \cite{dorfmann_magnetoelastic_2003} have set the foundations in this field by summarizing the Maxwell’s equations, mechanical balance laws, and thermodynamic equations for a deformable 3D continuum. This general theory was then specialized for magneto-sensitive elastomers, with simplified forms of the constitutive relations according to the incompressibility and hyperelasticity of the elastomer. Building up on the framework of~\cite{dorfmann_magnetoelastic_2003}, \cite{Zhao_JMPS2019} proposed a specific constitutive law for ideal hard-magnetic soft materials. The magnetic flux density induced in the material was assumed to be linear to the external field strength, with a constant of proportionality (permeability) close to the permeability of vacuum. \cite{Zhao_JMPS2019} implemented the equilibrium equation of a hard-magnetic body under a uniform field, along with the constitutive law, into 3D finite element modeling (FEM) through a user-defined element subroutine (UEL) developed in the commercial finite element software package Abaqus/Standard. The model was validated by experiments on a variety of 2D and 3D hard-magnetic structures.

In specific structural designs, a full 3D continuum theory is usually overkill and makes the analysis cumbersome. In comparison, reduced-order structural theories are desirable to understand the magnetic effect and further identify the key parameters of the system, in order to guide the design. For instance, beam-like hard-magnetic structural elements that are flexible and capable of large deflections in 3D space, have been used in many magneto-mechanical systems~\citep{Lum_PNAS2016,Hu_Nature2018,Kim_SciRobot2019,Gu_NatCommun2020,pancaldi_flow_2020,wang_evolutionary_2021}. The simple geometry of these 1D elements reduces the complexity in modeling, design, and fabrication. Based on the continuum theory discussed above for hard-MREs, several studies have investigated the mechanical response of thin beams to magnetic actuation, through a combination of modeling, simulation, and experiments~\citep{Lum_PNAS2016,Wang_JMPS2020,chen_complex_2020,chen_mechanics_2020,chen_theoretical_2020,chen_three-dimensional_2021,dehrouyeh-semnani_bifurcation_2021}. In Table~\ref{table:past_studies}, we summarize and classify these recent studies in terms of the methods used (rows) and the uniformity of applied fields (columns), indicating whether the effect of field gradients was considered.

In a pioneering study, \cite{Lum_PNAS2016} proposed a geometrically nonlinear inextensible beam model derived from force and torque balance of the beam centerline. The model was then used to optimize the magnetization profile of a beam to reach a target deformed configuration under either a uniform or a gradient field, and the results were validated experimentally. \cite{Wang_JMPS2020} proposed a hard-magnetic elastica theory for inextensible beams. The energy of the beam was written with respect to the centerline, and the governing equation was obtained through the principle of stationary action. The model was quantitatively examined under a uniform field by FEM simulations and experiments~\citep{Zhao_JMPS2019}. However, in the absence of a field gradient, the terms related to magnetic forces in the equilibrium equation, which were inconsistent with those reported in~\cite{Lum_PNAS2016}, were overlooked in the verification and validation of the framework. Further, to consider the stretch of the centerline, a beam model with the exact geometric nonlinearity under uniform fields was developed and used to predict the deformation of cantilever beams ~\citep{dehrouyeh-semnani_bifurcation_2021,chen_complex_2020,chen_mechanics_2020,chen_theoretical_2020}, finding negligible differences with the inextensible model. Taking the inextensible and the geometrically exact beam models, much effort has been focused on the (inverse) design of magnetization profiles~\citep{Lum_PNAS2016,ciambella_form-finding_2020, chen_complex_2020,chen_mechanics_2020,wang_evolutionary_2021} to optimize shape-shifting mode for specific applications.

\begingroup
\setlength{\tabcolsep}{8pt} 
\renewcommand{\arraystretch}{1.6} 
\begin{table}
\label{table:past_studies}
\small
\centering
   \caption{Past studies on hard-magnetic thin beams.}
   \begin{tabular}{  >{\centering\arraybackslash}p{1.9cm}   p{4.2cm}   p{4.2cm}  }
     \noalign{\hrule height 1.pt}
     \multicolumn{1}{c}{Past studies} &  \multicolumn{1}{c}{Uniform fields}  & \multicolumn{1}{c}{Gradient fields} \\ \noalign{\hrule height 1.pt}
     Theory &  
     \setlength{\parskip}{0.5em}
     \cite{Lum_PNAS2016,Wang_JMPS2020, wang_evolutionary_2021,chen_complex_2020,chen_mechanics_2020,chen_theoretical_2020,ciambella_form-finding_2020, dehrouyeh-semnani_bifurcation_2021} &
     
     \setlength{\parskip}{0.3em}
     \cite{Lum_PNAS2016,Wang_JMPS2020} \\ \hline
     Simulations &
     \setlength{\parskip}{0.3em}
     \cite{Zhao_JMPS2019,ye_magttice_2021}\\ \hline
     Experiments &
     \setlength{\parskip}{0.3em}
     \cite{Lum_PNAS2016,Wang_JMPS2020, wang_evolutionary_2021} & \cite{Lum_PNAS2016}\\ \noalign{\hrule height 1.pt}
   \end{tabular}
 \end{table}
 \endgroup

As conveyed by the relevant bibliographic summary in Table~\ref{table:past_studies}, to date, most of the research efforts on hard-magnetic beams have focused on uniform applied fields. However, spatially non-uniform fields are inevitably encountered when permanent magnets are used for actuation, for the purpose of miniaturization, movability, and the ease of manipulation ~\citep{Kim_SciRobot2019,xi_rolled-up_2013,tang_versatile_2018}. More importantly, field gradients (and the loads induced) have been shown to enhance the shape-shifting and locomotion performance~\citep{Lum_PNAS2016,Kim_SciRobot2019,Hu_Nature2018}. Also, due to the field heterogeneity, a body with less local constraints can have multiple stable configurations~\citep{Hu_Nature2018,zhang_dual-axial_2016}. To understand the effect of field gradients, there is a striking lack of studies on the deformation of thin beams subjected to combined magnetic torques and forces in non-uniform fields. In addition, the existing geometrically nonlinear beam model~\citep{Lum_PNAS2016,Wang_JMPS2020} was validated only in uniform fields. As such, there is a timely need for an experimental validation of the model in gradient fields, where magnetic body forces come into play. Likewise, the existing 3D FEM framework based on the continuum theory of bulk hard-magnetic solids is limited to uniform fields. This restriction calls for an effort to augment existing theories to also incorporate the effect of field gradients and implement them in FEM to yield a numerical tool that could simulate the behavior of hard-magnetic structures in any shape under non-uniform fields.

In this work, we provide a comprehensive predictive framework for the large deflections of hard-magnetic thin beams under uniform and gradient fields, comprising both a geometrically nonlinear inextensible beam model and 3D FEM. These two predictive tools are fully validated against precision experiments. Our FEM framework and experimental validation will address the missing gaps for gradient fields in Table~\ref{table:past_studies}. First, for the beam model, we start from the 3D elastic and magnetic energy and arrive at a 1D energy formulation through dimensional reduction~\citep{Audoly_ElasticityGeometry2010}. The equilibrium equation is obtained using variational methods. Although performing the reduction from 3D to 1D is cumbersome compared to directly examining the equilibrium or energy of the beam centerline~\citep{Lum_PNAS2016,Wang_JMPS2020}, this approach can serve as a starting point for modeling hard-magnetic structural elements with more complex geometries, such as rods~\citep{chen_three-dimensional_2021,sano_rods_2021}, plates, and shells~\citep{yan_magneto-active_2021}. In parallel to the dimensionally reduced theory, we extend the framework for 3D hard-magnetic solids under a uniform field proposed by~\cite{Zhao_JMPS2019} to the general case of an applied gradient field. In particular, the equilibrium equation of a 3D deformable body is derived from the principle of virtual work, by taking into account the magnetic body force imposed by the field gradient. Our extended theory is then implemented in FEM to verify the beam model. For the purpose of validation, we perform a series of precision experiments on cantilever beams, finding an excellent quantitative agreement between the model, FEM simulations, and experiments. To the best of our knowledge, this is the first time that the beam model for hard-MREs is verified and validated for the general case of a non-uniform magnetic field with gradients. Furthermore, using the experimentally validated framework, we are able to carry out comparative studies to investigate the mechanical response of a cantilever beam under different field configurations.

Our paper is organized as follows. Our problem is defined in Section~\ref{sec:problem}. In Section~\ref{sec:model}, we derive the hard-magnetic beam model by first reducing the energy from 3D to 1D and then obtaining the equilibrium equation through variational methods. We present the extended 3D theory of hard-magnetic solids in Section~\ref{sec:theory_FEM}. In Section~\ref{sec:EXP}, we describe our experimental methodology. The experimental results and the comparisons with the theory and FEM are presented in Section~\ref{sec:results}, followed by a series of comparative case studies in Section~\ref{sec:comparison}. Our findings are summarized and discussed in Section \ref{sec:conclusion}.

\section{Definition of the problem}
\label{sec:problem}
\begin{figure}[t!]
\centering
  \includegraphics[width=0.9\columnwidth]{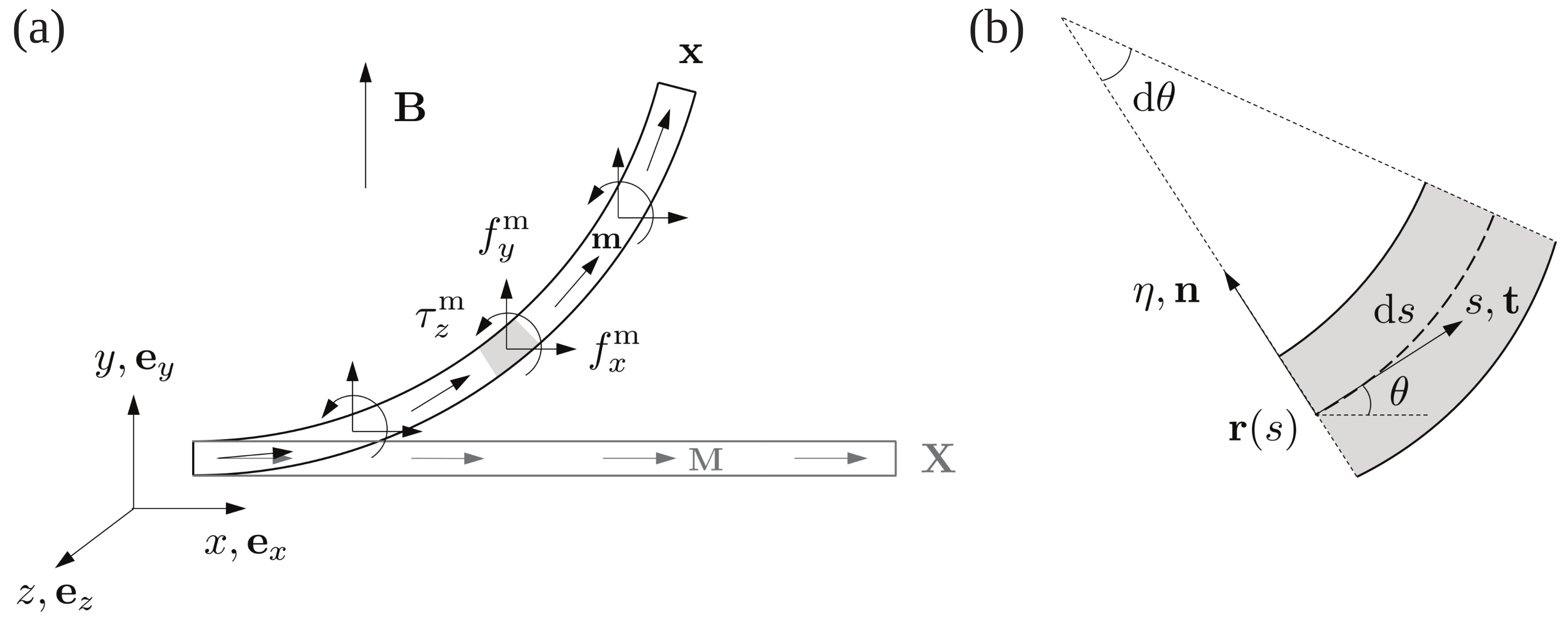}
  \caption{ (a) Schematic of a hard-magnetic thin beam under an external field. The beam deflects from the initial configuration $\mathbf{X}$ (magnetization $\mathbf{M}$) to the deformed configuration $\mathbf{x}$ (magnetization $\mathbf{m}$), subjected to the magnetic body torque ($\tau_{z}^\mathrm{m}$) and force (with components $f_{x}^\mathrm{m}$ and $f_{y}^\mathrm{m}$) imposed by the field (flux density $\mathbf{B}$). (b) Schematic of an infinitesimal segment of the beam, which is the magnification of the shaded region in (a). The centerline, parameterized by the arclength coordinate, $s$, is described using the tangent angle, $\theta$, between the tangent vector, $\mathbf{t}$, and the normal vector, $\mathbf{n}$.}
  \label{fig:Fig1}
\end{figure}

We consider a slender, straight, hard-magnetic beam, whose length, $L$, is much larger than its thickness, $h$, and width, $W$: $L/h,L/W \gg 1$ (see schematic in Fig.~\ref{fig:Fig1}). We parametrize the beam using the arc-length coordinate of its centerline $s\in [0,\,L]$ and the orthogonal coordinates, $\eta \in [-h/2,\,h/2]$ and $\xi \in [-W/2,\,W/2]$, that frame the cross-section. Then, the initial configuration of the beam can be described as  
\begin{equation}
\mathbf{X}=(s, \eta,\xi)\,,
\end{equation}
and the deformed configuration under external loading is denoted by $\mathbf{x}$.

The beam is made of a hard-magnetic material possessing a remnant magnetization denoted by $\mathbf{M}$, with respect to the initial configuration, $\mathbf{X}$. Since the beam is slender, we assume that the magnetization is constant across the cross-section but can be varied along the length direction as $\mathbf{M}(s)$. The material is isotropic and elastic, characterized by Young's modulus, $E$, and Poisson's ratio, $\nu$.

Magnetic loading is exerted on the beam by applying an external magnetic field. To take into account the effect of field heterogeneity in space, we superpose a gradient component onto a uniform field. In this work, we consider the specific case of a constant gradient field; a common configuration in applications~\citep{Lum_PNAS2016,kummer_octomag_2010,diller_six-degree--freedom_2016,Hu_Nature2018}. Thus, the flux density vector of the field at a material point in the deformed beam can be expressed as
\begin{equation}
\mathbf{B}(\mathbf{x})=\mathbf{B}^\mathrm{C}+(\nabla\mathbf{B})\left(\mathbf{x}-\mathbf{x}_\mathrm{p}\right)\,,
\label{B}
\end{equation}
where $\mathbf{B}^\mathrm{C}$ is the homogeneous part, $\nabla\mathbf{B}$ is the field gradient (constant), and $\mathbf{x}_\mathrm{p}$ is the position vector of the field origin. Then, by setting the gradient $\nabla\mathbf{B}=0$, the field reduces to a uniform field.

We seek to investigate the deformation of the beam under the magnetic field in Eq.~\eqref{B}. Our study will combine experiments, theoretical analyses, and simulations. We restrict our analyses on 2D in-plane ($x$-$y$) deflections. As such, any component of the magnetization and the external field vectors that could enable out-of-plane ($z$) deflections must vanish,
\begin{equation}
\mathbf{M}(s)=(\hat{M}_{\alpha},0)\,,
\label{M}
\end{equation}
\begin{equation}
\mathbf{B}^\mathrm{C}=(\hat{B}^\mathrm{C}_{\alpha},0)\,,
\label{BC}
\end{equation}
and 
\begin{equation}
\nabla\mathbf{B}=\begin{pmatrix}
    \nabla\hat{B}_{\alpha\beta} & \begin{matrix} 0 \\ 0 \end{matrix} \\
    \begin{matrix} 0 & 0 \end{matrix} & \!\!\nabla{B}_{zz}
\end{pmatrix}\,,
\label{dB}
\end{equation}
where hatted quantities denote in-plane components, with Greek indices $\alpha,\beta=1,2$ representing the two spatial coordinates. According to the Maxwell equations~\citep{Miyazaki_Ferromagnetism2012}, in the absence of electric current and time-varying electric fields, the magnetic field must satisfy
\begin{equation}
\nabla\cdot\mathbf{B}=0\,,
\end{equation}
\begin{equation}
\nabla\times\mathbf{B}=0\,,
\end{equation}
which, specialized to the field described in Eqs.~\eqref{B}, \eqref{BC} and~\eqref{dB}, read
\begin{equation}
\nabla\hat{B}_{\alpha\alpha}+\nabla{B}_{zz}=0\,,
\label{Gauss_law}
\end{equation}
\begin{equation}
\nabla{\hat{\mathbf{B}}}=\nabla{\hat{\mathbf{B}}}^{\mathrm{T}}\,.
\end{equation}
We note that, for cases involving out-of-plane deflections, a 3D rod theory~\citep{chen_three-dimensional_2021} is required, which is beyond the scope of the presented work, but has recently been addressed elsewhere~\citep{sano_rods_2021}.

\section{A geometrically nonlinear hard-magnetic beam model}
\label{sec:model}
In this section, we derive a reduced-order model to describe the large deflections of hard-magnetic beams subjected to an external magnetic field. First, we review the geometrically nonlinear kinematics of inextensible thin beams under the Kirchhoff assumption: normals to the beam centerline remain normal and unstretched during deformations. Then, we develop a 1D beam model by following a dimensional reduction procedure~\citep{Audoly_ElasticityGeometry2010} based on the 3D Helmholtz free energy proposed by~\cite{Zhao_JMPS2019} for ideal hard-magnetic soft materials. In particular, we reduce the 3D magnetic potential by integrating it over the cross-section of the beam, arriving at a 1D formulation. The bending (elastic) energy of the beam is given by Euler’s elastica~\citep{Audoly_ElasticityGeometry2010}. 
Finally, by enforcing that the variation of the total potential energy of the system (magnetic and elastic components) vanishes at equilibrium, we obtain the governing equation of the beam with a boundary term.

\subsection{Kinematics}
\label{sec:kinematics}

To obtain the deformation gradient of the beam for computing the magnetic potential, for completeness and to set the notation, we first present the classic kinematic description for thin elastic beams undergoing large deflections. Since the beam is slender, the deformation is dominated by bending, with negligible shear, and the following classic assumptions are appropriate: 
\begin{enumerate}[label=(\roman*)]
\item Kirchhoff assumption: The material lines normal to the beam centreline in the initial configuration remain normal and unstretched in the deformed configuration; 
\item Inextensiblity: The beam centerline is bent without stretching, as is assumed in Euler’s elastica~\citep{Audoly_ElasticityGeometry2010}.
\end{enumerate}
Under the above assumptions, the deformed configuration of the beam can be written as
\begin{equation}
\mathbf{x}=\mathbf{r}(s)+\eta\mathbf{n}(s)+\xi \mathbf{e}_z\,,
\end{equation}
where $\mathbf{r}(s)$ is the deformed beam centerline, and its unit normal vector in the $x$-$y$ plane is $\mathbf{n}(s)$. Using the tangential angle, $\theta(s)$, between the tangent vector $\mathbf{t}(s)$ of the centerline and the Cartesian base vector $\mathbf{e}_x$, we have
\begin{equation}
\mathbf{r}=\left(x,y,0\right)=\left(\int_0^s{\cos\theta(s')\mathrm{d}s'},\int_0^s{\sin\theta(s')\mathrm{d}s'}, 0\right)\,,
\label{deformed_configuration}
\end{equation}
and 
\begin{equation}
\mathbf{n}=(-\sin\theta,\cos\theta,0)\,.
\end{equation}

By definition, the deformation gradient at a material point ($s, \eta, \xi$) in the deformed configuration $\mathbf{x}$ with respect to the initial configuration $\mathbf{X}$ is 
\begin{equation}
\mathbf{F}=\frac{\partial\mathbf{x}}{\partial\mathbf{X}}=
\begin{pmatrix}
    \hat{R}_{\alpha\beta} & \begin{matrix} 0 \\ 0 \end{matrix} \\
    \begin{matrix} 0 & 0 \end{matrix} & \!\!1
\end{pmatrix}-\eta\theta '\begin{pmatrix}
    \hat{Q}_{\alpha\beta} & \begin{matrix} 0 \\ 0 \end{matrix} \\
    \begin{matrix} 0 & 0 \end{matrix} & \!\!0
\end{pmatrix}\,,
\label{F}
\end{equation}
where prime denotes differentiation with respect to~$s$, and
\begin{equation}
\hat{\mathbf{R}}=\begin{pmatrix}
\cos\theta & -\sin\theta\\
\sin\theta & \cos\theta
\end{pmatrix}\,,
\end{equation}
\begin{equation}
\hat{\mathbf{Q}}=\begin{pmatrix}
\cos\theta & 0\\
\sin\theta & 0
\end{pmatrix}\,.
\end{equation}
The matrix $\hat{\mathbf{R}}$ represents the rotation of the beam centerline, and $\eta\theta '\hat{\mathbf{Q}}$ relates to the stretch of material fibers in the axial direction, which is linear to the beam thickness and vanishes at the centerline. The kinematics reviewed in this section is classic, well-established, and valid for thin beams undergoing large deflections.

\subsection{Dimensional reduction of the energy: from 3D to 1D}
\label{sec:Um}

The Helmholtz free energy of hard-magnetic materials can be decomposed into an elastic part associated with mechanical deformations and a magnetic part arsing from the interactions between remnant magnetization and the external field~\citep{Zhao_JMPS2019}. Based on this decomposition, the total energy of a hard-magnetic beam is the summation of the bending energy, ${U}^\mathrm{b}$, and the magnetic potential, ${U}^\mathrm{m}$:
\begin{equation}
U={U}^\mathrm{b}+{U}^\mathrm{m}\,.
\label{total_dU}
\end{equation}
Next, we derive the 1D formulations of ${U}^\mathrm{b}$ and ${U}^\mathrm{m}$ through dimensional reduction. For beams subjected to other forms of loading, \textit{e.g.}, gravitational force (which is non-negligible in our experiments; see Section~\ref{sec:results}), the corresponding potentials should be added into~Eq.\eqref{total_dU} and reduced from 3D to 1D, following a similar procedure to that described next.

\paragraph{Magnetic potential} The material of the beam is assumed to be hard-magnetic, magnetized upon saturation under a strong applied field. Following~\cite{Zhao_JMPS2019}, the magnetic potential of hard-magnetic materials can be expressed using the deformation gradient given in Eq.~\eqref{F} and the external flux density and remnant magnetization defined in Eqs.~\eqref{B} and~\eqref{M}, respectively, as
\begin{equation}
\begin{split}
{U}^\mathrm{m}&=\int_{V}-\mathbf{F}\mathbf{M}\cdot\mathbf{B}\,\mathrm{d}V\\
&=\int^{L}_{0}\int^{\frac{W}{2}}_{-\frac{W}{2}}\int^{\frac{h}{2}}_{-\frac{h}{2}}\bigg\{{-{\hat{\mathbf{R}}\hat{\mathbf{M}}}\cdot\hat{\mathbf{B}}(\hat{\mathbf{r}})}\\
&-\eta{\big[\hat{\mathbf{R}}\hat{\mathbf{M}}}\cdot(\nabla\hat{\mathbf{B}}\hat{\mathbf{n}})-{\theta '\hat{\mathbf{Q}}\hat{\mathbf{M}}\cdot\hat{\mathbf{B}}(\hat{\mathbf{r}})\big]}\\
&+\eta^2{\theta '\hat{\mathbf{Q}}\hat{\mathbf{M}}\cdot(\nabla\hat{\mathbf{B}}\hat{\mathbf{n}})}\bigg\}\,\mathrm{d}\eta\mathrm{d}\xi\mathrm{d}s\,.
\end{split}
\label{Um_3D_beam}
\end{equation}
This 3D magnetic potential represents the work required to align the magnetization of the beam with the applied field.

The integrand in Eq.~\eqref{Um_3D_beam} contains terms at different orders in $\eta$. Recall that hatted quantities are defined in the $x-y$ plane. We reduce this energy potential by integration along the thickness ($\eta$) and the width ($\xi$) directions. The linear term in $\eta$ vanishes upon integration from $\eta=-h/2$ to $\eta=h/2$. The second-order term would give a bending-like energy, coexiting with the field gradient. This high-order term is neglected in the following derivation, given the excellent agreement that we achieved between the beam model derived by retaining only the zeroth order term, 3D FEM, and experiments (see Section~\ref{sec:results}). Therefore, Eq.~\eqref{Um_3D_beam} gives a stretching-like energy:
\begin{equation}
{U}^\mathrm{m}=-A\int_0^L{{\hat{\mathbf{R}}\hat{\mathbf{M}}}\cdot\hat{\mathbf{B}}(\hat{\mathbf{r}})}\,\mathrm{d}s\,,
\label{Um_1D}
\end{equation}
where $A=hW$ is the cross-sectional area of the beam. So far, the 3D magnetic potential has been reduced to a 1D form written on the beam centerline, akin to Euler’s elastica~\citep{Audoly_ElasticityGeometry2010}.

\paragraph{Elastic energy} Next, we consider the elastic energy of the beam, which has long been well studied in Euler's elastica~\citep{Audoly_ElasticityGeometry2010}. The stretching energy vanishes due to the assumption of inextensibility. The bending energy is given in terms of the curvature of the beam centerline,  $\kappa=\theta '$, as  
\begin{equation}
{U}^\mathrm{b}=\frac{1}{2}\int_0^L EI_\xi \theta '^2\mathrm{d}s\,,
\label{Ub_1D}
\end{equation}
where $I_\xi=Wt^3/12$ is the area moment of inertia of the beam's cross-section with respect to the $\xi$-axis.

\subsection{Equilibrium equation}

Thus far, we have provided the 1D formulations for the magnetic potential, Eq.~\eqref{Um_1D}, and the classic bending energy, Eq.~\eqref{Ub_1D}, of hard-magnetic thin beams. We proceed to derive the equilibrium equation through the calculus of variations. According to the condition of stationary, the first variation of the energy functional of the system must varnish for any arbitrary perturbation $\delta\theta$,
\begin{equation}
\delta U=U(\theta+\delta\theta)-U(\theta)=\delta {U}^\mathrm{b}+\delta {U}^\mathrm{m}=0\,.
\label{dU}
\end{equation}
Next, we will compute each term of the first variation of the total energy, individually (\textit{i.e.}, $\delta {U}^\mathrm{m}$ and $\delta {U}^\mathrm{b}$), and then obtain the equilibrium equation from Eq.~\eqref{dU}. 

\paragraph{The first variation of magnetic potential $\delta {U}^\mathrm{m}$} According to the reduced magnetic potential derived in Eq.~\eqref{Um_1D}, its first variation with respect to $\delta\theta$ is 
\begin{equation}
\delta{U}^\mathrm{m}=-A\int_0^L\left[{\hat{\mathbf{R}}_{,\theta}\hat{\mathbf{M}}}\cdot\hat{\mathbf{B}}(\hat{\mathbf{r}})\delta\theta +{\hat{\mathbf{R}}\hat{\mathbf{M}}}\cdot\left(\nabla\hat{\mathbf{B}}\begin{pmatrix}
\int_0^s{-\delta\theta\sin\theta\,\mathrm{d}s'}\\
\int_0^s{\delta\theta\cos\theta\,\mathrm{d}s'}
\end{pmatrix}
\right)\right]\,\mathrm{d}s\,,
\label{dUm_1}
\end{equation}
where~$(\cdot),_{\theta}$ denotes derivative with respect to~$\theta$.

Before we can obtain the equilibrium equation from Eq.~\eqref{dU}, we first need to rewrite the second term of the integrand of Eq.~\eqref{dUm_1} proportionally to $\delta \theta$. By changing the order of integration in the double integral, Eq.~\eqref{dUm_1} becomes  
\begin{equation}
\delta{U}^\mathrm{m}=-A\int_0^L\left[{{\hat{\mathbf{R}}_{,\theta}\hat{\mathbf{M}}}\cdot\hat{\mathbf{B}}(\hat{\mathbf{r}})}+ {\hat{\mathbf{n}} \cdot\int_s^L{\nabla\hat{\mathbf{B}}^{\mathrm{T}}(\hat{\mathbf{R}}\hat{\mathbf{M}})}}\,\mathrm{d}s'\right]\delta\theta\,\mathrm{d}s \,,
\label{dUm_2}
\end{equation}
where we recognize the magnetic body torque about the $z$-axis  
\begin{equation}
\hat{\mathbf{R}}_{,\theta}\hat{\mathbf{M}}\cdot\hat{\mathbf{B}}(\hat{\mathbf{r}})=[\hat{\mathbf{R}}{\mathbf{M}}\times\hat{\mathbf{B}}(\hat{\mathbf{r}})]\cdot{\mathbf{e}_z}=\tau_{z}^\mathrm{m}\,,
\label{m_torque}
\end{equation}
and the in-plane magnetic body force vector  
\begin{equation}
\nabla\hat{\mathbf{B}}^{\mathrm{T}}(\hat{\mathbf{R}}\hat{\mathbf{M}})=\hat{\mathbf{f}}^\mathrm{m}=({f}_{x}^\mathrm{m},{f}_{y}^\mathrm{m})\,.
\label{m_force}
\end{equation}
Combining all of the above results, the first variation of magnetic energy is 
\begin{equation}
\delta{U}^\mathrm{m}=-A\int_0^L{\left(\tau_{z}^\mathrm{m}
-\sin\theta\int_s^L{{f}_{x}^\mathrm{m}}\,\mathrm{d}s'
+\cos\theta\int_s^L{{f}_{y}^\mathrm{m}}\,\mathrm{d}s'\right)}\delta\theta\,\mathrm{d}s\,.
\label{dUm}
\end{equation}

\paragraph{The first variation of bending energy $\delta {U}^\mathrm{b}$} From the expression of ${U}^\mathrm{b}$ in Eq.~\eqref{Ub_1D}, we have
\begin{equation}
\delta{U}^\mathrm{b}=\int_0^L {EI_\xi}{\theta '\delta\theta '}\,\mathrm{d}s\,,
\label{dUb_1}
\end{equation}
which can be rewritten using integration by parts, as
\begin{equation}
\delta{U}^\mathrm{b}={EI_\xi}\theta '\delta\theta\,|_{0}^{L}-\int_0^L{EI_\xi}{\theta''}\,\delta\theta\mathrm{d}s\,.
\label{dUb}
\end{equation}

With the first variations of the individual energy terms in Eqs.~\eqref{dUm} and~\eqref{dUb}, we invoke the stationary condition of Eq.~\eqref{dU} and obtain the equilibrium equation
\begin{equation}
EI_\xi\theta''+A\tau_{z}^\mathrm{m}-A\sin\theta\int_s^L{{f}_{x}^\mathrm{m}}\,\mathrm{d}s'+A\cos\theta\int_s^L{{f}_{y}^\mathrm{m}}\,\mathrm{d}s'=0\, 
\label{Equi_Eq}
\end{equation}
for hard-magnetic beams under a gradient field, with a boundary term
\begin{equation}
\theta '\delta\theta\,|_{0}^{L}=0\,.
\label{Equi_BT}
\end{equation}

To help identify relevant parameters in specific problems, we provide the dimensionless form of Eq.~\eqref{Equi_Eq}
\begin{equation}
\theta_{,\bar{s}\bar{s}}+\overline{\tau}_{z}^\mathrm{m}-\sin\theta\int_{\bar{s}}^1{\overline{f}_{x}^\mathrm{m}}\,\mathrm{d}\bar{s}'+\cos\theta\int_{\bar{s}}^1{\overline{f}_{y}^\mathrm{m}}\,\mathrm{d}\bar{s}'=0\,,
\label{Equi_Eq_nondimensionless}
\end{equation}
and the boundary term
\begin{equation}
\theta_{,\overline{s}}\delta\theta\,|_{0}^{1}=0\,,
\label{Equi_BT_nondimensionless}
\end{equation}
where we have defined a dimensionless arc-length coordinate $\bar{s}={s}/{L}$, and $(),_{\bar{s}}$ denotes derivative with respect to~$\bar{s}$. The corresponding dimensionless forms of the magnetic body torque and force are
\begin{equation}
\overline{\tau}_{z}^\mathrm{m}=\frac{AL^2}{EI_\xi}{\tau}_{z}^\mathrm{m}\,
\label{m_torque_nondimensionless}
\end{equation}
and
\begin{equation}
\overline{\hat{\mathbf{f}}}^\mathrm{m}=\frac{AL^3}{EI_\xi}{\hat{\mathbf{f}}}^\mathrm{m}\,,
\label{m_force_nondimensionless}
\end{equation}
respectively.

We can see from Eq.~\eqref{Equi_Eq_nondimensionless} together with Eqs.~\eqref{m_torque} and~\eqref{m_force} that, under a constant gradient field, both the body torque and force are imposed on the beam. However, under a uniform field (\textit{i.e.}, $\hat{\mathbf{B}}=\hat{\mathbf{B}}^\mathrm{C}$ with $\nabla\hat{\mathbf{B}}=0$), the body force vanishes, $\overline{\hat{\mathbf{f}}}^\mathrm{m}=0$. Thus, the beam is subjected to only the body torque $\overline{\tau}_{z}^\mathrm{m}$, and Eqs.~\eqref{Equi_Eq_nondimensionless} and~\eqref{Equi_BT_nondimensionless} reduce to
\begin{equation}
\theta_{,\bar{s}\bar{s}}+\overline{\tau}_{z}^\mathrm{m}=0\,,
\end{equation}
\begin{equation}
\theta_{,\overline{s}}\delta\theta\,|_{0}^{1}=0\,.
\end{equation}

In this section, based on the Helmholtz free energy of hard-magnetic materials, we have derived a reduced-order model for thin hard-magnetic beams. By solving the boundary value problem defined in Eqs.~\eqref{Equi_Eq_nondimensionless} and~\eqref{Equi_BT_nondimensionless} with appropriate boundary conditions, we can predict the deformation of the beam under an applied magnetic field. Mechanical loads can be taken into account by adding the corresponding reduced potentials to the total energy of the system. In Section~\ref{sec:results}, we use this theory to analyze the deflection of a hard-magnetic cantilever beam in response to a uniform or a constant gradient field. The equilibrium equation is solved using the \textit{bvp5c} solver in Matlab under given boundary conditions. In Section~\ref{sec:results}, we will compare the predictions from this magnetic beam model against FEM (developed in Section~\ref{sec:theory_FEM}) and experiments (developed in Section~\ref{sec:EXP}).

We note that, the equilibrium equation, Eq.~\eqref{Equi_Eq_nondimensionless}, obtained above is identical to that derived by~\cite{Lum_PNAS2016} from force and torque balance of the beam centerline. Based on the 3D formulation of hard-magnetic solids, our approach provides a rigorous derivation for the beam model by taking advantage of dimensional reduction. In the future, this same approach can be extended to develop reduced models for other slender structural elements~\citep{yan_magneto-active_2021,sano_rods_2021}. Without following a dimensional reduction procedure, \cite{Wang_JMPS2020} directly wrote the 1D magnetic energy on the beam centerline, which is consistent with Eq.~\eqref{Um_1D}. However, a mistake was made in the calculus of variations, leading to an equilibrium equation, Eq.~(30) in \cite{Wang_JMPS2020}, that is erroneously inconsistent with our result and that reported in~\cite{Lum_PNAS2016}. These challenges provide an additional motivation for the more rigorous approach that we have followed toward founding a solid basis for the modeling of hard-MRE beams.

\section{Modeling of bulk hard-magnetic materials under gradient fields}
\label{sec:theory_FEM}

To verify the reduced beam model presented in the previous section, we now develop a full 3D theory for bulk hard-magnetic materials under gradient fields and implement this theory in FEM. Our work is based on the macroscopic continuum framework proposed recently by~\cite{Zhao_JMPS2019} for hard-magnetic materials under uniform fields. In Section~\ref{sec:theory_3D}, we extend this existing theory to incorporate the effect of field gradients, which impose a distributed body force in the material. Specifically, staring from the Helmholtz free energy of hard-magnetic solids, we invoke the principle of virtual work (PVW) to derive the corresponding equilibrium equation of a 3D elastic continuum body in the presence of field gradients. The magnetic part of the constitutive law of the material is determined, relating the magnetically-induced stresses to the magnetization of the deformed body and the applied field. Then, in Section~\ref{sec:FEM_3D}, the extended 3D theory is implemented in Abaqus/Standard through a user-defined element (subroutine UEL), which is modified from that developed in~\cite{Zhao_JMPS2019} (restricted to uniform fields) to also consider the case of field gradients. We provide the equilibrium equation at element's nodes and the Jacobian of element's residual, which are required when defining the user element. The numerical implementation through 3D FEM (together with the beam model) will be validated against experiments in Section~\ref{sec:results}.

\subsection{Equilibrium equation and constitutive relation}
\label{sec:theory_3D}

We consider a continuum body of volume $V$ bounded by surface $S$ in its current configuration. The elastic and hard-magnetic body is subjected to non-magnetic body forces (per unit volume), $\mathbf{f}$, and a gradient field, $\mathbf{B}$. The magnetic loading (torques and/or forces) exerted by the field will be considered by the magnetic potential of the body. At $S$, traction forces (per unit area), $\mathbf{t}$, and prescribed displacements, $\mathring{\mathbf{u}}$, are applied on complementary sub-surfaces, $S_t$ and $S_u$, respectively. 

The body deforms under the given loading and boundary conditions. At equilibrium, the first variation of the Helmholtz free energy vanishes for an arbitrary perturbation $\delta \mathbf{u}$ ($\delta\mathbf{u}=0$ on $S_u$ to satisfy boundary conditions):
\begin{equation}
\delta{U}=\delta{U}^\mathrm{e}+\delta{U}^\mathrm{m}-\int_{S_t}{\mathbf{t}\cdot\delta\mathbf{u}}\,\mathrm{d}S-\int_V{\mathbf{f}\cdot\delta\mathbf{u}}\,\mathrm{d}V=0\,,
\label{variation_3D}
\end{equation}
where ${U}^\mathrm{e}$ is the elastic energy of the material, determined by the constitutive relation in the absence of magnetic fields (\textit{e.g.}, a neo-Hookean model). By definition, the first variation of elastic energy~\citep{Gurtin_MechanicsThermodynamicsContinua2010,bower_applied_2009,ogden_non-linear_1997} is 
\begin{equation}
\delta{U}^\mathrm{e}=\int_V{\boldsymbol{\sigma}^\mathrm{e}:\frac{\partial\delta\mathbf{u}}{\partial\mathbf{x}}}\,\mathrm{d}V\,,
\label{dUe_3D}
\end{equation}
where $\boldsymbol{\sigma}^\mathrm{e}$ is the elastic part of Cauchy stress.

The magnetic energy of the body is  ${U}^\mathrm{m}=\int_{V}-\mathbf{m}\cdot\mathbf{B}\,\mathrm{d}V$~\citep{Zhao_JMPS2019}, where $\mathbf{m}=J^{-1}\mathbf{F}\mathbf{M}$ is the remnant magnetization in the current configuration. We compute the first variation of ${U}^\mathrm{m}$ as 
\begin{equation}
\begin{split}
\delta{U}^\mathrm{m}&=\int_V\big[{-J^{-1}(\delta\mathbf{F})\mathbf{M}\cdot\mathbf{B}-\mathbf{m}\cdot\delta\mathbf{B}}\big]\,\mathrm{d}V \\
&=\int_V\big[{-\left(\mathbf{B}\otimes\mathbf{m}\right):\frac{\partial\delta\mathbf{u}}{\partial\mathbf{x}}-\left(\nabla\mathbf{B}^{\mathrm{T}}\mathbf{m}\right)\cdot\delta\mathbf{u}}\big]\,\mathrm{d}V\,,
\end{split}
\label{dUm_3D}
\end{equation}
where $\delta\mathbf{B}=\nabla\mathbf{B}\delta\mathbf{u}$ for a constant gradient field. We note that, the vector field $\mathbf{B}$ in Eq.~\eqref{dUm_3D} denotes the applied flux density in the region occupied by the deformed body; \textit{i.e.}, in a heterogeneous field, it depends on the motion of the body. As such, we retain the term containing $\delta\mathbf{B}$ in Eq.~\eqref{dUm_3D}, which turns out to be the source of magnetic body force. 

Then, we invoke PVW of a hard-magnetic elastic body to obtain the equilibrium equation by substituting Eqs.~\eqref{dUe_3D} and~\eqref{dUm_3D} into Eq.~\eqref{variation_3D}, as
\begin{equation}
\int_V{\left(\boldsymbol{\sigma}^\mathrm{e}-\mathbf{B}\otimes\mathbf{m}\right):\frac{\partial\delta\mathbf{u}}{\partial\mathbf{x}}}\,\mathrm{d}V =\int_{S_t}{\mathbf{t}\cdot\delta\mathbf{u}}\,\mathrm{d}S+\int_V{\left(\mathbf{f}+\nabla\mathbf{B}^{\mathrm{T}}\mathbf{m}\right)\cdot\delta\mathbf{u}}\,\mathrm{d}V\,.
\label{virtual_work_1}
\end{equation}
From the external virtual work, the right hand side of Eq.~\eqref{virtual_work_1}, we indentify the magnetic body force
\begin{equation}
\mathbf{f}^\mathrm{m}=\nabla\mathbf{B}^{\mathrm{T}}\mathbf{m}\,. 
\label{magnetic_body_force}
\end{equation}
The stress-like term introduced by the field in the internal virtual work, the left hand side of Eq.~\eqref{virtual_work_1}, defines the magnetic part of Cauchy stress:
\begin{equation}
\boldsymbol{\sigma}^\mathrm{m}=-\mathbf{m}\otimes\mathbf{B}\,.
\label{magnetic_stress_3D}
\end{equation}
Eq.~\eqref{magnetic_stress_3D} describes the magnetic contribution to the constitutive relation of hard-magnetic materials. Therefore, the total Cauchy stress is
\begin{equation}
\boldsymbol{\sigma}=\boldsymbol{\sigma}^\mathrm{e}+\boldsymbol{\sigma}^\mathrm{m}\,,
\label{Cauchy_stress_3D}
\end{equation}
which is asymmetric due to the magnetic part, $\boldsymbol{\sigma}^\mathrm{m}$. The corresponding first Piola-Kirchhoff stress~\citep{ogden_non-linear_1997} is
\begin{equation}
\mathbf{P}=J\boldsymbol{\sigma}^{\mathrm{T}}\mathbf{F}^{\mathrm{-T}}=\mathbf{P}^\mathrm{e}+\mathbf{P}^\mathrm{m}\,,
\label{stress_P-K}
\end{equation}
where $\mathbf{P}^\mathrm{e}=J\boldsymbol{\sigma}^{\mathrm{e}}\mathbf{F}^{\mathrm{-T}}$ and $\mathbf{P}^\mathrm{m}=-\mathbf{B}\otimes\mathbf{M}$ are the elastic and magnetic parts, respectively. Finally, considering Eqs.~\eqref{magnetic_body_force} and~\eqref{Cauchy_stress_3D}, the PVW in Eq.~\eqref{virtual_work_1} is rewritten as
\begin{equation}
\int_V{\boldsymbol{\sigma}^\mathrm{T}:\frac{\partial\delta\mathbf{u}}{\partial\mathbf{x}}}\,\mathrm{d}V =\int_{S_t}{\mathbf{t}\cdot\delta\mathbf{u}}\,\mathrm{d}S+\int_V{\left(\mathbf{f}+\mathbf{f}^\mathrm{m}\right)\cdot\delta\mathbf{u}}\,\mathrm{d}V\,.
\label{virtual_work_2}
\end{equation}

Thus far, we have invoked the PVW for a hard-magnetic elastic body, from which the magnetic part of constitutive relation and the magnetic body force were determined. We proceed to derive the equilibrium equation of the body by performing an integration by parts and applying the Gauss' theorem on the left hand side of Eq.~\eqref{virtual_work_2}, to obtain
\begin{equation}
\int_{S_t}{\mathbf{n}\cdot\boldsymbol{\sigma}\cdot\delta\mathbf{u}}\,\mathrm{d}S -\int_V{\frac{\partial}{\partial\mathbf{x}}\cdot\boldsymbol{\sigma}\cdot\delta\mathbf{u}}\,\mathrm{d}V=\int_{S_t}{\mathbf{t}\cdot\delta\mathbf{u}}\,\mathrm{d}S+\int_V{\left(\mathbf{f}+\mathbf{f}^\mathrm{m}\right)\cdot\delta\mathbf{u}}\,\mathrm{d}V\,,
\label{equilibrium_1}
\end{equation}
where $\mathbf{n}$ is the normal vector of the surface ${S_t}$. Eq.~\eqref{equilibrium_1} is preserved for any arbitrary $\delta\mathbf{u}$ that is kinematically compatible. Consequently, the equilibrium equation is
\begin{equation}
\frac{\partial}{\partial\mathbf{x}}\cdot\boldsymbol{\sigma}+\mathbf{f}+\mathbf{f}^\mathrm{m}=0\,,
\label{equilibrium_2}
\end{equation}
where the total Cauchy stress $\boldsymbol{\sigma}=\boldsymbol{\sigma}^\mathrm{e}+\boldsymbol{\sigma}^\mathrm{m}$, and the boundary conditions are
\begin{equation}
\mathbf{t}=\mathbf{n}\cdot\boldsymbol{\sigma}\,\,\,\, \mathrm{on}\,\,\,\,S_t\,,\,\,\,\mathbf{u}=\mathring{\mathbf{u}}\,\,\,\, \mathrm{on}\,\,\,\,S_u\,.
\label{equilibrium_BC}
\end{equation}
One can verify that the balance of angular momentum~\citep{dorfmann_magnetoelastic_2003,Zhao_JMPS2019}
\begin{equation}
\varepsilon:\left(\frac{\boldsymbol{\sigma}-\boldsymbol{\sigma}^\mathrm{T}}{2}\right)+\boldsymbol{\tau}=0\,
\end{equation}
where $\boldsymbol{\tau}=\mathbf{m}\times\mathbf{B}$ is the magnetic body torque imposed by the field and $\varepsilon$ is the third-order permutation tensor, is naturally satisfied by the Cauchy stress in Eq.~\eqref{Cauchy_stress_3D}.

In this section, thus far, starting from the Helmholtz free energy of hard-magnetic materials, we have obtained the equilibrium equation, Eq.~\eqref{equilibrium_2}, of a 3D elastic body under gradient fields, by invoking the PVW, Eq.~\eqref{virtual_work_1}. The magnetic effect includes a contribution to the stress and a body force. The equations are given in the current (deformed) configuration, and the corresponding expressions with respect to the initial configuration can be derived by following the same procedure.

Comparing the 3D theory presented above with that proposed by~\cite{Zhao_JMPS2019} for the case of uniform fields, the magnetic part of the first Piola-Kirchhoff stress in Eq.~\eqref{stress_P-K} is a dependent variable on the body's motion, due to the non-uniformity of the field. Moreover, a magnetic body force, $\mathbf{f}^\mathrm{m}$, is imposed on the material, which indeed appears in the equilibrium equation, Eq.~\eqref{equilibrium_2}. \cite{Zhao_JMPS2019} and~\cite{Wang_JMPS2020,wang_evolutionary_2021} stated that the magnetic body force should not be incorporated into the equilibrium equation as a body force term, since the magnetic effect has all been taken into account by the total stress tensor containing a magnetic part. However, this is not the case, as shown by the rigorous derivation performed above that will be validated by experiments in Section~\ref{sec:results}. Our result is also in accordance with the general framework proposed by \cite{dorfmann_magnetoelastic_2003} for magneto-sensitive elastic solids. At last, we note that the magnetic part of Cauchy stress in Eq.~\eqref{magnetic_stress_3D} is the transpose of that given by~\cite{Zhao_JMPS2019}. This is because we followed the convention that the traction force on a surface of normal $\mathbf{n}$ is $\mathbf{t}=\mathbf{n}\cdot\boldsymbol{\sigma}$, Eq.~\eqref{equilibrium_BC} in this paper, rather than $\mathbf{t}=\boldsymbol{\sigma}\mathbf{n}$, Eq.~(2.12) in \cite{Zhao_JMPS2019}. The two expressions are different when the stress tensor is asymmetric, as is the case considered in the present study.

\subsection{Numerical implementation in FEM}
\label{sec:FEM_3D}

We solve the boundary value problem defined by Eqs.~\eqref{equilibrium_2} and~\eqref{equilibrium_BC} using the general finite element method, which enables analyses of the mechanical response to constant gradient fields of 3D hard-magnetic structures in general geometries. The numerical implementation is realized in the commercial finite element software package Abaqus/Standard through a user-defined element, which is modified from the element developed by~\cite{Zhao_JMPS2019} for simulations under uniform fields. Specifically, according to the theory proposed in Section~\ref{sec:theory_3D}, the magnetic body force induced by the field gradient should be considered in the element's equilibrium, and the Jacobian of element's residual has to be re-derived since the flux density applied on the element now depends on the deformed configuration. We will describe our numerical implementation with a focus on the required modifications to the original pioneering work of~\cite{Zhao_JMPS2019}, which, nonetheless, we still follow closely.

The implementation in FEM is based on the PVW for a hard-magnetic body in Eq.~\eqref{virtual_work_2}, following a standard discretization procedure. Then, the weak form at the element level (volume $V^\mathrm{element}$, surface $S_t^\mathrm{element}$ subjected to traction forces $\mathbf{t}$) can be expressed as
\begin{equation}
\int_{V^\mathrm{element}}\boldsymbol{\sigma}^\mathrm{T}:\frac{\partial\delta\mathbf{v}}{\partial\mathbf{x}}\,\mathrm{d}V=\int_{S_t^\mathrm{element}}{\mathbf{t}\cdot\delta\mathbf{v}}\,\mathrm{d}S+\int_{V^\mathrm{element}}{\left(\mathbf{f}+\mathbf{f}^\mathrm{m}\right)\cdot\delta\mathbf{v}}\,\mathrm{d}V\,,
\label{weak_form}
\end{equation}
where $\mathbf{u}=\sum_{I=1}^{n}\mathbf{u}^{I}{N}^{I}$ and $\delta\mathbf{v}=\sum_{I=1}^{n}\delta\mathbf{v}^{I}{N}^{I}$ are, respectively, approximated displacement field and virtual velocity field, interpolated from the corresponding nodal quantities, $\mathbf{u}^{I}$ and $\delta\mathbf{v}^{I}$, using the same shape function $N^{I}$ (node number $I=1,2,...n$). The element's residual at the nodes for a intermediate solution is
\begin{equation}
\mathbf{R_u}^{I}=-\int_{{V}^{\mathrm{element}}}\boldsymbol{\sigma}^\mathrm{T}\frac{\partial{N}^{I}}{\partial\mathbf{x}}\,\mathrm{d}V+\int_{{V}^{\mathrm{element}}}{N}^{I}\left(\mathbf{f}+\mathbf{f}^\mathrm{m}\right)\,\mathrm{d}V+\int_{S_t^\mathrm{element}}{N}^{I}{\mathbf{t}}\,\mathrm{d}S\,.
\label{Ru_FEM}
\end{equation}
Therefore, the nodal displacements at equilibrium can be obtained by solving $\mathbf{R_u}^{I}=0$ using the Newton–Raphson method, requiring the Jacobian of element's residual to update the intermediate solution in each iteration; \textit{i.e.},
\begin{equation}
\mathbf{K_{uu}}^{IJ}=-\frac{\partial\mathbf{R_u}^{I}}{\partial\mathbf{u}^{J}}\,.
\label{Kuu_1}
\end{equation}
where $I,J=1,2,...n$ are node numbers in an element.

Given a certain form of elastic energy ${U}^\mathrm{e}$, one can derive the expression of Cauchy stress $\boldsymbol{\sigma}$ using Eqs.~\eqref{dUe_3D}, \eqref{magnetic_stress_3D}, and \eqref{Cauchy_stress_3D} and compute the Jacobian of element's residual, $\mathbf{K_{uu}}^{IJ}$, according to Eq.~\eqref{Kuu_1}. In our simulations, we choose the neo-Hookean model~\citep{bower_applied_2009} to describe the elastic behavior of the hard-magnetic elastomer used in our experiments (the experimental details and procedures will be provided in Section~\ref{sec:EXP}). The expressions of the elastic parts of stress tensors, $\mathbf{P}^\mathrm{e}$ and $\boldsymbol{\sigma}^\mathrm{e}$, for neo-Hookean materials can be found in~\cite{Zhao_JMPS2019}. Here, we provide the expression of $\mathbf{K_{uu}}^{IJ}$ in the absence of traction forces $\mathbf{t}$ and non-magnetic body forces $\mathbf{f}$, which is specific for constant gradient fields: 
\begin{equation}
\begin{split}
\left(\mathbf{K_{uu}}^{IJ}\right)_{ij}=\int_{{V}^{\mathrm{element}}}\bigg(&J^{-1}\frac{\partial{N}^{I}}{\partial x_k}\frac{\partial{P^\mathrm{e}_{im}}}{\partial{F_{jn}}}F_{km}F_{ln}\frac{\partial{N}^{J}}{\partial x_l}\\
&-J^{-1}\frac{\partial{N}^{I}}{\partial x_k}F_{kl}{M}_l\nabla{B}_{ij}{N}^{J} \\
&-J^{-1}{N}^{I}F_{kl}{M}_l\nabla{B}_{ij}\frac{\partial{N^J}}{\partial x_k}\bigg)\,\mathrm{d}V\,.
\end{split}
\label{Kuu_2}
\end{equation}
where $i,j=1,2,3$ denote the $i$-th and $j$-th spatial coordinates. The last two terms in Eq.~\eqref{Kuu_2} are contributions from the field gradient $\nabla{\mathbf{B}}$, which would vanish in the case of uniform fields. 

To avoid the volumetric locking problem when modeling incompressible materials, \cite{Zhao_JMPS2019} used the F-Bar method~\citep{de_souza_neto_design_1996}. In this method, the deformation gradient $\mathbf{F}$ at a material point is modified as 
\begin{equation}
\overline{\mathbf{F}}:=\left(\frac{\mathrm{det}(\mathbf{F}_0)}{\mathrm{det}(\mathbf{F})}\right)^{\frac{1}{3}}\mathbf{F}\,,
\end{equation}
where $\mathbf{F}_0$ is the deformation gradient at the element centroid. Then, the determinant of $\overline{\mathbf{F}}$ is equal to that of the element centroid, \textit{i.e.}, $J=\mathrm{det}(\overline{\mathbf{F}})=\mathrm{det}(\mathbf{F}_0)$, and the point-wise constraint of incompressibility in an element is relaxed to a single constraint at the centroid.

Under the scheme of the F-Bar method, the residual nodal forces in Eq.~\eqref{Ru_FEM} is computed by simply replacing $\mathbf{F}$ with $\overline{\mathbf{F}}$, and the corresponding Jacobian of element's residual is 
\begin{equation}
\begin{split}
\left(\mathbf{K_{uu}}^{IJ}\right)_{ij}=\int_{{V}^{\mathrm{element}}}\bigg[&J^{-1}\frac{\partial{N}^{I}}{\partial x_k}\frac{\partial{P^\mathrm{e}_{im}}}{\partial{\overline{F}_{jn}}}\overline{F}_{km}\overline{F}_{ln}\frac{\partial{N}^{J}}{\partial x_l}\\
&\frac{1}{3}J^{-1}\frac{\partial{N}^{I}}{\partial x_k}\frac{\partial{P^\mathrm{e}_{im}}}{\partial{F_{ln}}}\overline{F}_{km}\overline{F}_{ln}\left(\frac{\partial{N^J}}{\partial x_j}-\frac{\partial{N^J}}{\partial x_j}|_0\right)\\
&-\frac{2}{3}\frac{\partial{N}^{I}}{\partial x_k}\sigma_{ki}\left(\frac{\partial{N^J}}{\partial x_j}-\frac{\partial{N^J}}{\partial x_j}|_0\right) \\
&-J^{-1}\frac{\partial{N}^{I}}{\partial x_k}\overline{F}_{kl}{M}_l\nabla{B}_{ij}{N}^{J} \\
&-J^{-1}{N}^{I}\overline{F}_{kl}{M}_l\nabla{B}_{ij}\frac{\partial{N^J}}{\partial x_k}\\
&-\frac{2}{3}{N}^{I}{f}^\mathrm{m}_i\left(\frac{\partial{N^J}}{\partial x_j}-\frac{\partial{N^J}}{\partial x_j}|_0\right)\bigg]\,\mathrm{d}V\,,
\end{split}
\label{Kuu_3}
\end{equation}
where $(\cdot)|_0$ denotes quantities evaluated at the element centroid.

The above element description under gradient fields is applied to a user-defined eight-node hexahedron element, which was originally developed by~\cite{Zhao_JMPS2019} for FEM under uniform fields. We have modified the element's residual and its Jacobian, according to Eqs.~\eqref{Ru_FEM} and~\eqref{Kuu_3}, to be able to simulate the behavior of hard-magnetic structures subjected to gradient fields.

In particular, we have conducted 3D FEM simulations on thin beams to verify the proposed reduced model (Section~\ref{sec:model}), and both the 3D simulations and the reduced-order theory will be validated through experiments (Section~\ref{sec:results}). In the simulations, a cantilever beam was discretized by a regular mesh with $10\times10$ elements in the cross-section. The element size in the length direction was determined in such a way that the aspect ratios of the elements were close to 1. We performed a convergence study to ensure that the mesh was sufficiently fine. Geometric nonlinearities were taken into consideration throughout the simulations (an option in Abaqus/Standard). During loading, a magnetic field was applied on the beam in the same configuration as that of the experiments (see Section~\ref{sec:EXP}). We also considered gravitational forces, which were imposed on the beam through a congruent dummy mesh. The geometric, elastic, and magnetic (input) parameters used in the simulations were experimentally measured, independently, such that there were no fitting parameters. 

\section{Experiments}
\label{sec:EXP}

In this section, we present the experimental methodology followed to study the deformation of hard-magnetic beams under an applied field. First, we will describe the protocols that we have developed to prepare the MRE material (Section~\ref{sec:MRE}) and fabricate the beam specimens (Section~\ref{sec:fab_beams}). Then, in Section~\ref{sec:fields}, we elaborate on the generation of either a uniform or a gradient magnetic field. The experimental configurations, parameters, and procedures are described in Section~\ref{sec:exp_protocol}.

\subsection{Preparation of the MRE}
\label{sec:MRE}

The MRE used in our experiments was a composite of Vinylpolysiloxane polymer (VPS-22, Elite Double, Zhermack) embedded with NdPrFeB particles (hard-magnetic, average diameter of $5\,\mu$m, MQFP-15-7-20065-089, Magnequench). The initially liquid MRE mixture used to fabricate beam specimens upon curing (Section~\ref{sec:fab_beams}) was prepared in the following three steps: 

\begin{enumerate}[label=(\roman*)]

\item \textit{Mixing}. The non-magnetized NdPrFeB particles were added into the liquid VPS-22 base, with a mass ratio of 2:1. This suspension was mixed using a centrifuge (ARE-250, Thinky Corporation) for $40\,$s at $2000\,$rpm (mixing mode) and another $20\,$s at $2200\,$rpm (defoaming mode).

\item \textit{Degassing}. The solution was degassed in a vacuum chamber (absolute pressure below $8\,$mbar) to eliminate air bubbles trapped during the initial mixing process.

\item \textit{Adding catalyst}. The same amount of VPS-22 catalyst to that of the VPS-22 base was added into the mixture obtained during step (ii). After another mixing step for $20\,$s at $2000\,$rpm (mixing mode), followed by $10\,$s at $2200\,$rpm (defoaming mode), the liquid MRE was ready to be used for the fabrication of beam specimens, which cured in 15-20 min.

\end{enumerate}

The final fraction of NdPrFeB particles in the MRE was $50.0\%$ in mass. The densities of the VPS-22 matrix and NdPrFeB particles were, respectively, $1.16\,\mathrm{g/cm^3}$ (measured using a pycnometer) and $7.61\,\mathrm{g/cm^3}$ (provided by the supplier). According to the masses and densities of the two phases, we computed the volume fraction of the particles in our MRE to be $13.2\%$. The density of the MRE was $\rho=2.01\pm0.05\,\mathrm{g/cm^3}$, given by the law of mixtures~\citep{alger_polymer_2017}.

We characterized the Young's modulus of the cured MRE through cantilever tests. Three beams of width $3.36\pm0.54$\,mm were cut off from a plate of thickness $2.420\pm0.012$\,mm cast using the previously prepared liquid MRE. In the tests, each beam was clamped at three different positions to vary its effective length between $36$\,mm and $50$\,mm. At each length, the cantilever beam deflected under self-weight and its shape was captured by a camera. The unknown Young's modulus of the MRE was determined by minimizing the difference between the shape of the deformed beam given by the Euler's elastica \citep{Audoly_ElasticityGeometry2010} and that measured in the experiments. The tests on the three specimens at three different lengths for each specimen yielded an average Young's modulus of $E=1.16\pm0.04\,$MPa. The Poisson's ratio of the MRE was assumed to be $\nu\approx0.5$ due to the incompressibility of elastomers.

\subsection{Fabrication and magnetization of beam specimens} 
\label{sec:fab_beams}

\begin{figure}[t!]
\centering
  \includegraphics[width=1\columnwidth]{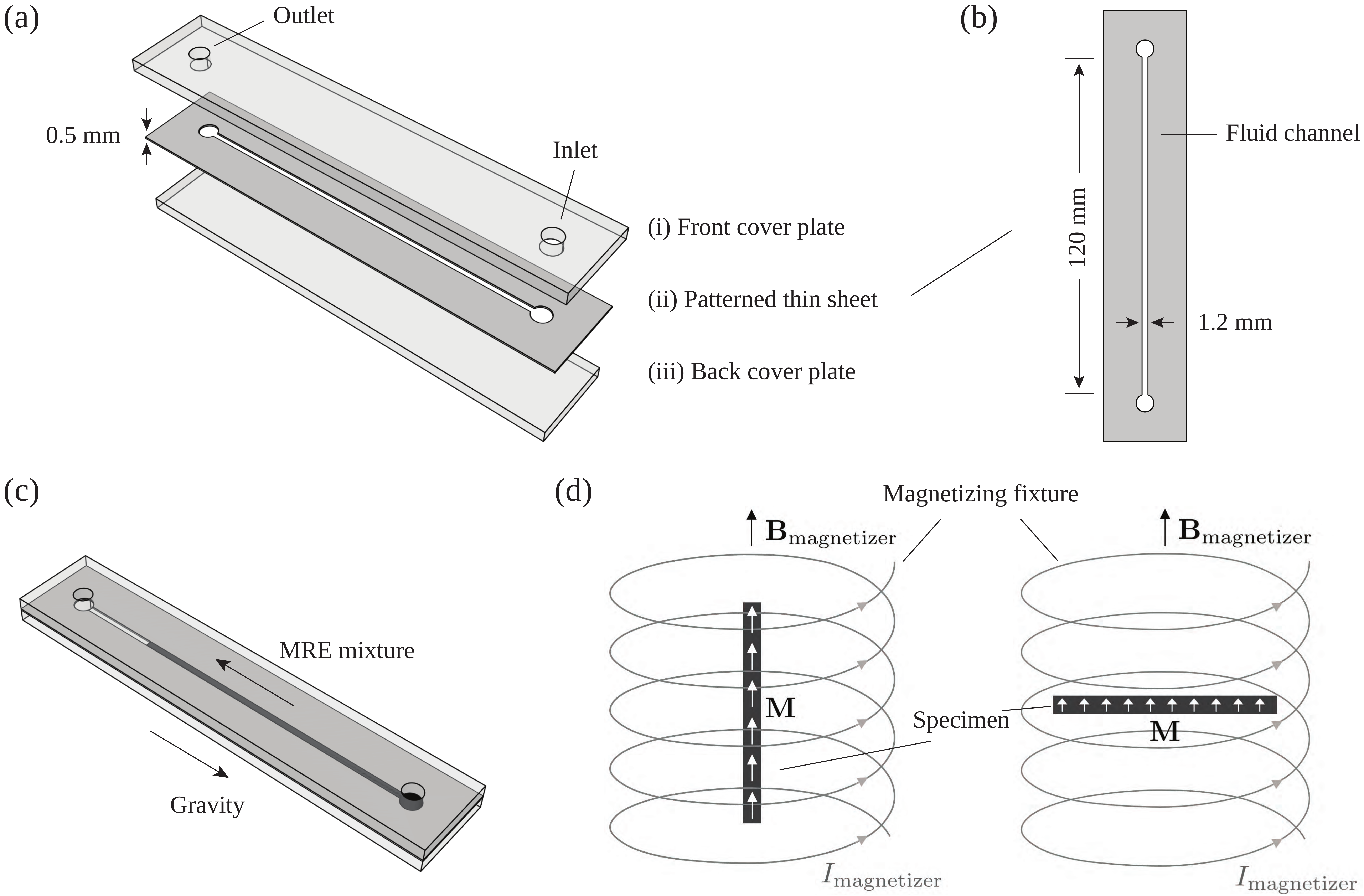}
  \caption{Schematics of the fabrication and magnetization of beam specimens. (a) The sandwich structure of the mold: (i) a front cover plate, (ii) a patterned thin sheet, and (iii) a back cover plate. (b) Dimensions of the channel laser-cut in the thin sheet to cast the beams. (c) Liquid MRE is injected from the bottom inlet to fill up the mold. (d) After curing, the beam specimen is magnetized in an electromagnetic coil (magnetizing fixture). The current $I_\mathrm{magnetizer}$ provided by a pulse magnetizer generates a strong axial magnetic field $B_\mathrm{magnetizer}\approx4.4\,T$ that saturates the particles. The beam can be either placed vertically or horizontally to realize magnetization in the length (left) or thickness (right) direction, respectively.}
  \label{fig:Fig2}
\end{figure}

As shown schematically in Fig.~\ref{fig:Fig2}(a), the mold used for the casting of the beam specimens was a sandwich structure consisted of (i) a front cover plate, (ii) a patterned thin sheet, and (iii) a back cover plate. The sheet was made from a plastic shim stock of thickness $0.5$\,mm (The Artus Corporation) and perforated with a narrow channel of dimensions $120$\,mm length and  $2$\,mm width using a last cutter (see Fig.~\ref{fig:Fig2}b).

During beam fabrication, the prepared MRE solution was injected into the mold through the \textit{inlet} (in the front plate) using a syringe (see schematic in Fig.~\ref{fig:Fig2}c), until it filled up the channel. The mold was placed vertically throughout the fabrication process, such that the air inside the channel was expelled through the \textit{outlet}. Upon curing, we first magnetized the beam specimen by saturating the embedded hard-magnetic particles using a pulse magnetizer (IM-K-010020-A, flux density $\approx4.4$\,T, Magnet-Physik Dr. Steingroever GmbH), and the beam was then removed form the mold. Fig.~\ref{fig:Fig2}(d) shows the direction of magnetization, which could be oriented either in the beam's length or thickness direction.

The saturated hard-magnetic particles possess a remnant flux density of $0.90\,$T (reported by the supplier). We assume that the particles were randomly distributed in the polymer matrix with negligible re-arrangements during magnetizing. As such, the composite could be considered homogenized as a continuum solid with a uniform magnetization profile. Then, the magnitude of magnetization of our MRE was computed to be the volume-average of the total magnetic moment of individual particles; \textit{i.e.}, ${M} = 94.1\,$kA/m.

\subsection{Generation of either a uniform or a gradient field}
\label{sec:fields}

Next, we focus on the generation of the applied magnetic field. We considered two representative configurations for the magnetic field, when studying the deformation of our beam specimens: either (i) a uniform or (ii) a constant gradient field. 

The magnetic fields were generated by two identical multi-turn electromagnetic coils with inner and outer radii of $36\,$mm and $57\,$mm, respectively (average radius $R_\mathrm{c}=46.5$\,mm).
The two coils were set concentrically and powered by a DC power supply providing a maximum current/power of $25\,$A/$1.5\,$kW (EA-PSI 9200-25 T, EA-Elektro-Automatik GmbH). The centre-to-centre distance and polarities of the two coils were adjusted to set either the Helmholtz configuration or the Maxwell configuration to generate, respectively, a uniform or a constant gradient field, as detailed next.

\begin{figure}[t!]
\centering
  \includegraphics[width=1\columnwidth]{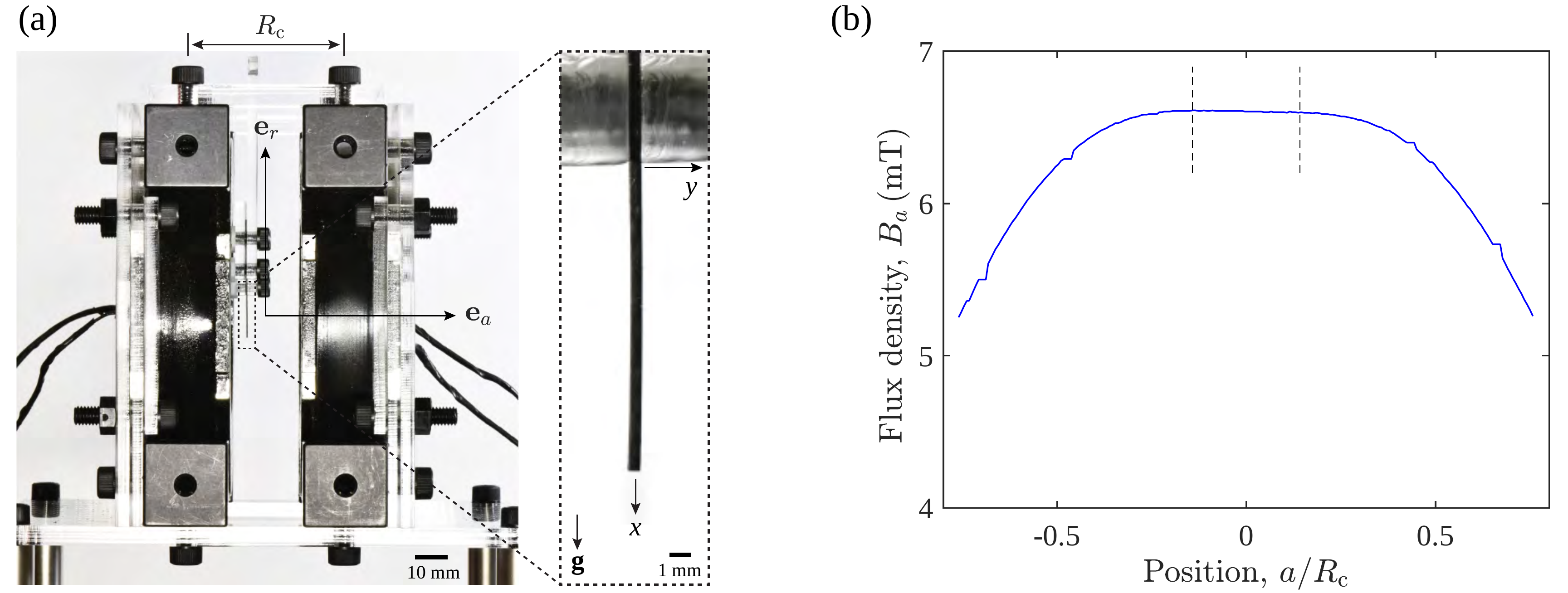}
  \caption{Generation of a uniform magnetic field using Helmholtz coils. (a) Photograph of the coils set with a center-to-center distance of $R_\mathrm{c}$. A beam specimen was suspended in the central region with its top end clamped between two acrylic plates (see the magnified inset with the image of the beam). (b) The spatial variation of flux density $B_{a}$ in the axial direction ($\mathbf{e}_a$) of the two coils was measured by a Teslameter at $I_\mathrm{c}=1\,$A. The two dash lines indicate the region reached by the deformed beam in our experiments, where the field is uniform: $0.14 \leq a/R_\mathrm{c} \leq 0.14$.} 
  \label{fig:Fig3} 
\end{figure}

\paragraph{A uniform magnetic field} In the Helmholtz configuration (see Fig.~\ref{fig:Fig3}a), the two coils were separated with a centre-to-centre distance of $R_\mathrm{c}$. The currents, $I_\mathrm{c}$, in the two coils flowed in the same direction, thus generating two identical magnetic fields, whose superposition yielded a uniform axial field in the central region between the two coils. The field homogeneity was characterized by measuring the flux density along the axis of the two coils using a Teslameter (FH 55, Magnet-Physik Dr. Steingroever GmbH) at $I_\mathrm{c}=1\,$A, as shown in Fig.~\ref{fig:Fig3}(b). In the central region, in between the dashed lines in the plot ($0.14 \leq a/R_\mathrm{c} \leq 0.14$), the field vector is well described by 
\begin{equation}
\mathbf{B}=(0,{B}_{a},0)\,,
\label{BC_exp}
\end{equation}
and the non-zero component in the axial direction ($\mathbf{e}_a$), ${B}_{a}= 6.6\,\mathrm{mT}$, is constant in space and linear to $I_\mathrm{c}$. The field polarity could be flipped by reversing the currents in the two coils. In this configuration, the field gradient vanishes; $\nabla\mathbf{B}=0$.

\begin{figure}[t!]
\centering
  \includegraphics[width=1\columnwidth]{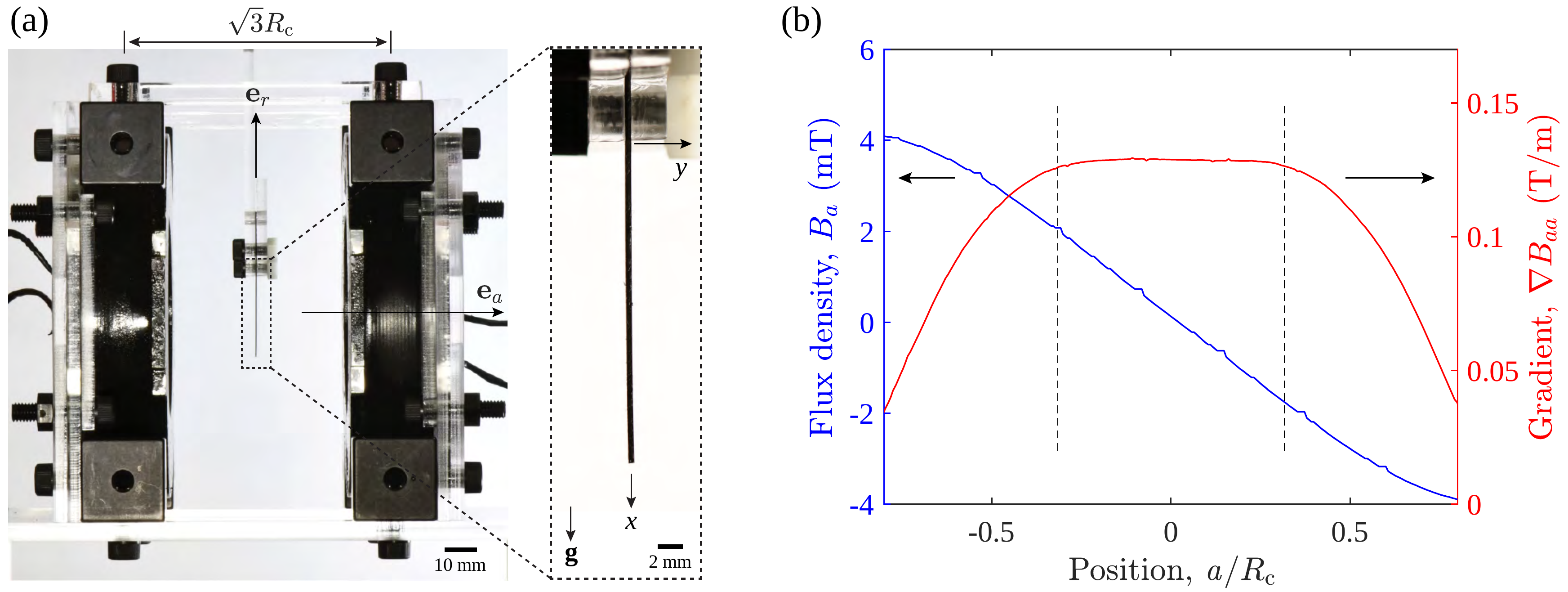}
  \caption{Generation of a constant gradient magnetic field using Maxwell coils. (a) Photograph of the coils set with a center-to-center distance of $\sqrt{3}R_\mathrm{c}$. A beam specimen was suspended in the central region with its top end clamped between two acrylic plates (see the magnified inset with the image of the beam). (b) The axial component of flux density vector, $B_{a}$ (measured by a Teslameter at $I_\mathrm{c}=1\,$A), and its gradient, $\nabla B_{aa}$ (computed through numerical differentiation), along the axis of the coils ($a/R_\mathrm{c}$). The two dash lines indicate the region reached by the deformed beam in our experiments, where the field gradient is constant: $0.32 \leq a/R_\mathrm{c} \leq 0.32$.}
  \label{fig:Fig4}
\end{figure}

\paragraph{A constant gradient field} In the Maxwell configuration (see Fig.~\ref{fig:Fig4}a), the distance between the two coils was adjusted to $\sqrt{3}R_\mathrm{c}$, and the currents in the coils flowed in opposite directions. In the central region, the gradient of the superposed field is well described, in the Cartesian coordinate system, by 
\begin{equation}
\nabla\mathbf{B}=\begin{pmatrix}
\nabla B_{rr}  & 0  & 0\\
0 & \nabla B_{aa} & 0 \\
0 & 0 & \nabla B_{rr} 
\end{pmatrix}\,,
\label{dBG_exp}
\end{equation}
where $\nabla B_{aa}$ and $\nabla B_{rr}$ are the gradients of $B_{a}$ and $B_\mathrm{r}$ in the axial ($\mathbf{e}_a$) and radial ($\mathbf{e}_r$) directions, respectively.

In Fig.~\ref{fig:Fig4}(b), we plot the measured axial flux density, ${B}_{a}$, along the axis of the two coils at current $I_\mathrm{c}=1\,$A, which was zero at the center point. Its gradient in the axial direction is given by the numerical differentiation of ${B}_{a}$, which yields $\nabla{B}_{aa}= 0.13\,$T/m in the central region, in between the dashed lines in the plot ($0.32 \leq a/R_\mathrm{c} \leq 0.32$). For the Maxwell configuration, according to Eq.~\eqref{Gauss_law} and considering the axisymmetry of the field about $\mathbf{e}_a$, we have
\begin{equation}
\nabla{B}_{rr}= -\frac{1}{2}\nabla{B}_{aa}\,.
\end{equation}

\subsection{Experimental configurations, parameters, and procedures}
\label{sec:exp_protocol}

We experimentally studied the mechanical response of our beam specimens under the generated uniform and gradient magnetic fields. Before we present the results in Section~\ref{sec:results}, we first describe the configurations of the fabricated specimens, parameters explored, and the procedures followed to quantify the deformation of the beam.

The beam specimens for experiments under the uniform field had dimensions of thickness $t=0.50\pm0.01\,$mm, length $L=13.2\pm0.2\,$mm, and width $W=1.27\pm0.07\,$mm (average values and standard deviations on three specimens, labeled by s1-1, s1-2, and s1-3). They were magnetized in the length direction, as
\begin{equation}
\mathbf{M}=({M},0,0)\,.
\label{M_BC}
\end{equation}
The experiments were performed on the three, otherwise identical, beam specimens to examine uncertainty and reproducibility.

In the initial configuration shown in Fig.~\ref{fig:Fig3}(a), the field generated by the Helmholtz coils, Eq.~\eqref{BC_exp}, was applied perpendicularly to the magnetization of the beam, $\mathbf{M}$, Eq.~\eqref{M_BC}. As a result, a magnetic body torque was imposed over the entire beam, and the initially straight beam deformed into a curved configuration. We gradually increased the applied flux density in the range $0 \le B_{a} \le 66$\,mT by increasing the current $0 \le I_\mathrm{c} \le 10$\,A. At each increment, the rest configuration of the beam was imaged by a camera, and the shape of the centerline and the corresponding displacements were acquired using digital image processing (MATLAB). We note that, throughout the experiments, the beam was manipulated within the uniform region of the field (bounded by the dashed lines in Fig.~\ref{fig:Fig3}b), $0.14 \leq a/R_\mathrm{c} \leq 0.14$. 

Under the constant gradient field, in order to reach deep into the nonlinear regime, we fabricated another three specimens, labeled by s2-1, s2-2, and s2-3, with thickness $t=0.49\pm0.01\,$mm, length $L=25.8\pm0.3\,$mm, and width $W=1.21\pm0.05$\,mm (average values and standard deviations on the three specimens), to attain a higher slenderness ratio. Fig.~\ref{fig:Fig4}(a) shows the initial configuration of the beam. Considering the inhomogeneity of the field, we suspended the beam accurately in the symmetry plane between the two coils, and the center of the beam was aligned with the center of the field. Then, according to Eqs.~\eqref{B} and~\eqref{dBG_exp}, the field applied on the beam was
\begin{equation}
\mathbf{B}(\mathbf{x})=(\nabla\mathbf{B})(\mathbf{x}-\mathbf{x}_\mathrm{p})\,,
\label{BG_rp_exp}
\end{equation}
with $\mathbf{x}_\mathrm{p}=(L/2,0,0)$. The magnetization vector was perpendicular to the beam centerline
\begin{equation}
\mathbf{M}=(0,{M},0)\,.
\label{M_BG}
\end{equation}
In this constant gradient field configuration, the beam deformed in the $x-y$ plane subjected to combined magnetic body torques and forces. We adjusted the applied current $0 \le I_\mathrm{c} \le 12$\,A to generate a field gradient in the range $0 \le \nabla B_{aa} \le 1.56$\,T/m. The deflections and angles were measured from the deformed shapes of the beam imaged by a camera. 

\section{Results: Deformation of hard-magnetic beams under external fields}
\label{sec:results}
The proposed theoretical (Section~\ref{sec:model}), computational (Section~\ref{sec:theory_FEM}), and experimental (Section~\ref{sec:EXP}) methodologies enable us to study the deformation of hard-magnetic thin beams under uniform and gradient fields. In this section, we analyze the magnetic loading and the response of the beam in both of these field configurations. The beam model will be first verified through 3D FEM simulations, and both the theoretical and numerical predictions are then compared with experimental results for validation. For the sake of completeness and comparison with the case of a gradient field, we include the case of a uniform field that had already been studied by~\cite{Wang_JMPS2020} and~\cite{Zhao_JMPS2019}. The identified governing parameters and the quantitative results serve in helping us gain insight into the underlying magneto-elastic coupling effect.

\subsection{Hard-magnetic beam under a uniform magnetic field} 
\label{sec:constant_field}

We start with the configuration of a uniform field.
The specialization of Eq.~\eqref{m_torque_nondimensionless} for the magnetization profile of the beam, Eq.~\eqref{M_BC}, and the uniform field applied in our experiments, Eq.~\eqref{BC_exp}, gives the dimensionless magnetic body torque imposed on the beam
\begin{equation}
\overline{\tau}_{z}^\mathrm{m}=\frac{AL^2}{I_\xi}\frac{MB_a}{E}\cos{\theta}\,.
\label{magnetic_torque_BC}
\end{equation}
In this case, since the field is uniform (without a gradient, $\nabla\mathbf{B}=0$), the magnetic body force, Eq.~\eqref{m_force_nondimensionless}, vanishes:
\begin{equation}
\overline{\hat{\mathbf{f}}}^\mathrm{m}=0\,.
\label{magnetic_force_BC}
\end{equation}
Hence, the beam deformation is driven by the magnetic body torque alone. In addition, we take into account the effect of self-weight, which cannot be neglected in our experiments. The dimensionless gravitational force exerted on the beam is~\citep{Audoly_ElasticityGeometry2010}
\begin{equation}
\overline{\hat{\mathbf{f}}}^\mathrm{g}=\left(\overline{f}^\mathrm{g}_x,\overline{f}^\mathrm{g}_y\right)=\frac{AL^3\rho g}{EI_\xi} \mathbf{e}_x\,,
\label{g_force_nondimensionless}
\end{equation}
where is $\rho$ the density of the material and $g$ is the gravitational acceleration. 

Substituting Eqs.~\eqref{magnetic_torque_BC} and~\eqref{magnetic_force_BC} into Eq.~\eqref{Equi_Eq_nondimensionless} and incorporating the gravitational term~\citep{Audoly_ElasticityGeometry2010}, we obtain the equilibrium equation of the beam 
\begin{equation}
\theta_{,\bar{s}\bar{s}}+\lambda_\mathrm{m}^\mathrm{C}\cos{\theta}-(1-\bar{s})\overline{f}^\mathrm{g}_x\sin\theta=0\,.
\label{beam_uniform}
\end{equation}
The dimensionless magneto-elastic parameter
\begin{equation}
\lambda_\mathrm{m}^\mathrm{C}=\frac{MB_{a}AL^2}{EI_\xi}
\label{lambda_uniform}
\end{equation}
characterizes the relative importance of magnetic loading and beam bending effects.
We also notice that the governing equation is identical to that of a cantilever beam with a (dimensionless) point load of magnitude $\lambda_\mathrm{m}^\mathrm{C}$ applied at the tip~\citep{Audoly_ElasticityGeometry2010}, indicating that the magnetic body torque is equivalent to a mechanical point load in terms of the reaction forces and moments generated in the beam.

To solve Eq.~\eqref{beam_uniform}, we need appropriate boundary conditions, which, for a cantilever beam fixed at one end ($\bar{s}=0$) with the other end ($\bar{s}=1$) free, are
\begin{equation}
\theta(0)=0\,,\,\,\,\theta_{,\overline{s}}(1)=0\,.
\label{BC_uniform}
\end{equation}
One can verify that the boundary term in Eq.~\eqref{Equi_BT_nondimensionless} is satisfied by Eq.~\eqref{BC_uniform}.

\begin{figure}[t!]
\centering
  \includegraphics[width=1\columnwidth]{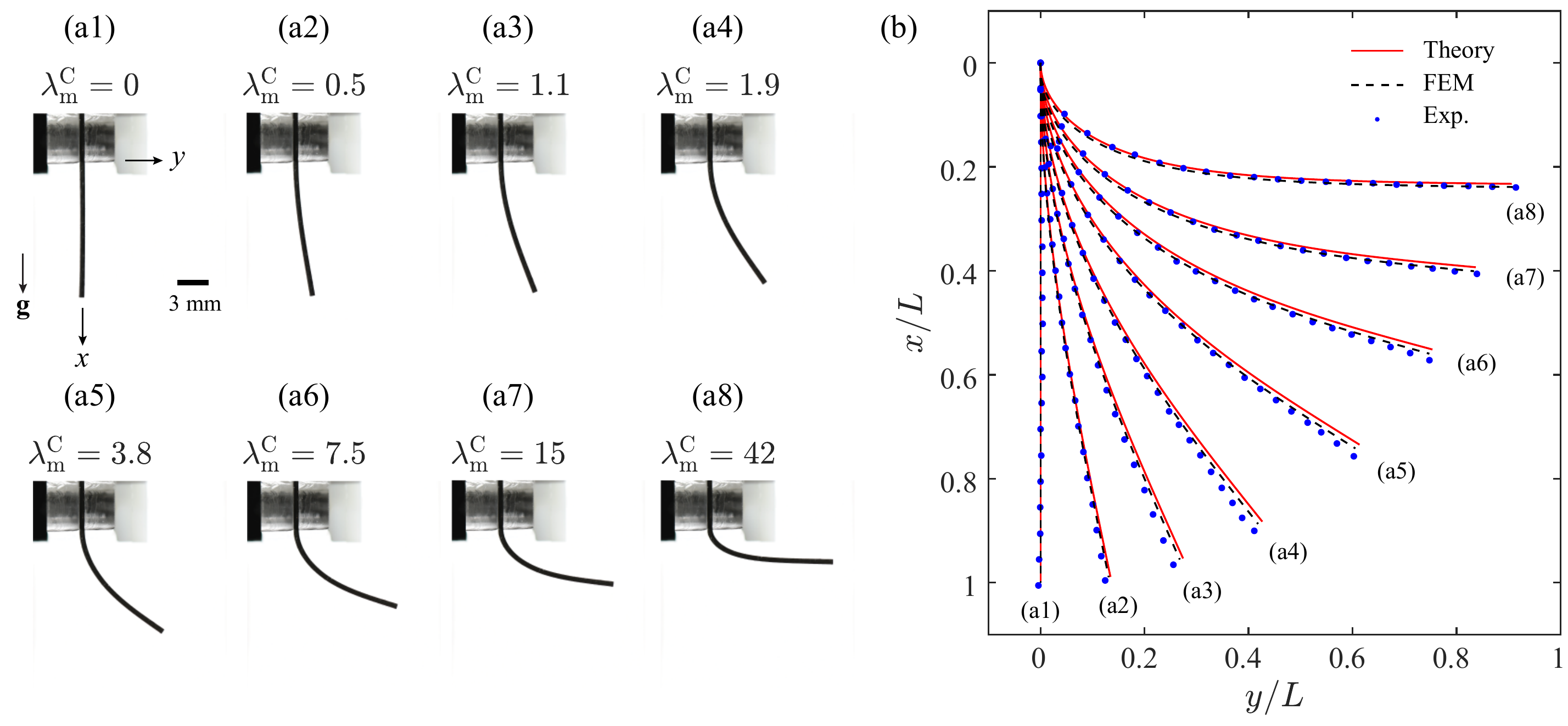}
  \caption{Configurations of a hard-magnetic beam subjected to a uniform magnetic field and gravity. The beam is magnetized in the $x$ direction in the initial configuration, perpendicular to the applied field in the $y$ direction (Section~\ref{sec:exp_protocol}). The gravitational field is in the $x$ direction. (a1)-(a8) Experimental photographs of the deformed beam at different levels of field strength, quantified by the parameter $\lambda_\mathrm{m}^\mathrm{C}$; the respective values are indicated above each photograph. (b) Beam centerlines (data points) exacted from the photographs in (a), together with the corresponding predictions by the beam model (solid lines) and FEM (dashed lines). }
  \label{fig:Fig5}
\end{figure}

Next, we quantify the deformation of the beam in the experiments and contrast the results against the theoretical and numerical predictions. Fig.~\ref{fig:Fig5}(a1)-(a8) presents experimental photographs of the deformed beam at different levels of magnetic loading ($\lambda_\mathrm{m}^\mathrm{C}$). The centerlines of the beam, extracted from the photographs using image processing, are shown in Fig.~\ref{fig:Fig5}(b). We observed that the deflection increases with the field strength and the beam undergoes large rotations in the range of the field explored. At the extreme case of $\lambda_\mathrm{m}^\mathrm{C}=42$ in Fig.~\ref{fig:Fig5}(a8), the rotation angle at the tip reaches almost $90^\circ$.

In parallel to the experiments, we performed FEM simulations to compute the deformation of the beam using the 3D element proposed in Section~\ref{sec:FEM_3D}. In Fig.~\ref{fig:Fig5}(b), the predictions from FEM and the specialized beam model, Eq.~\eqref{beam_uniform}, at the 8 loading levels are superimposed on the experimental results. Excellent agreement is found between our beam model, FEM and experiments, thereby providing a first step of validation of our predictive framework. We re-emphasize that all the parameters used in the theory and FEM were measured independently in the experiments (Section~\ref{sec:EXP}), with no fitting parameters.

\begin{figure}[t!]
\centering
  \includegraphics[width=1\columnwidth]{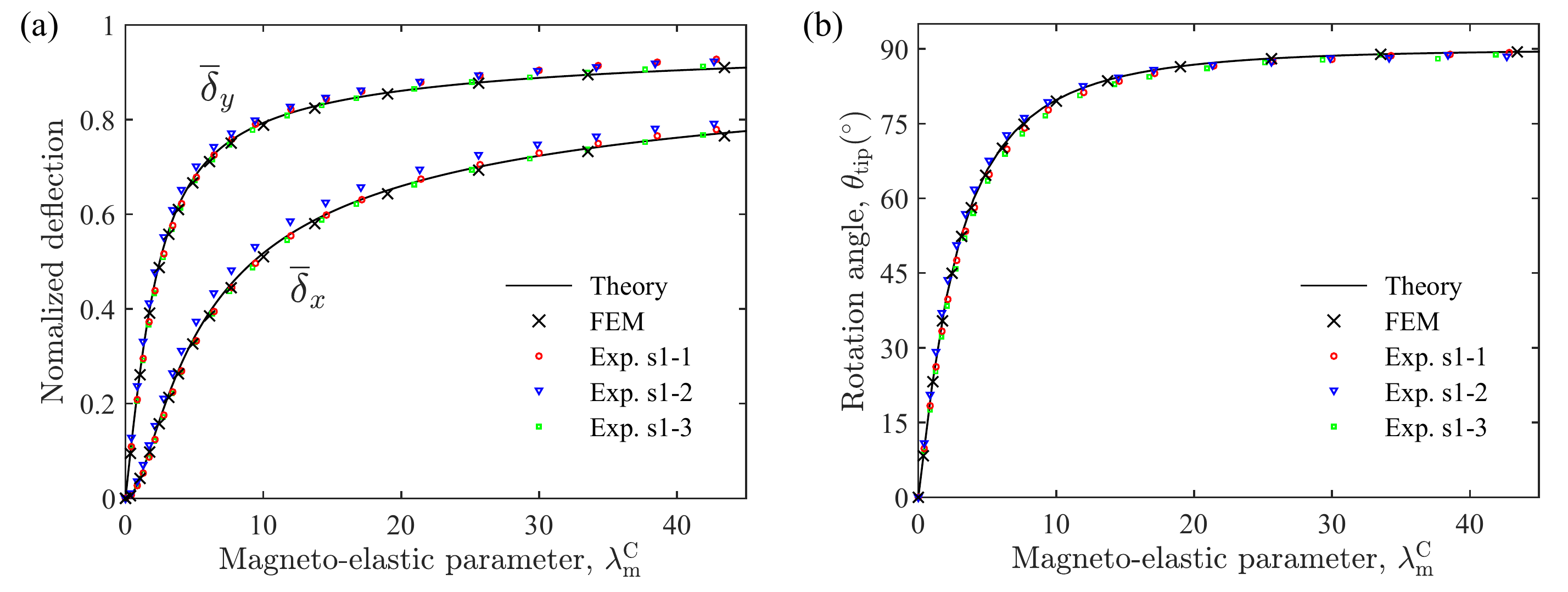}
  \caption{Beam deformation under a uniform magnetic field. (a) Normalized components of the deflection, $\overline{\delta}_x=|\delta_x|/L$ (absolute value) and $\overline{\delta}_y=\delta_y/L$, and (b) rotation angle, $\theta_\mathrm{tip}$, at the free end of the deformed beam versus the dimensionless magneto-elastic parameter $\lambda_\mathrm{m}^\mathrm{C}$. The results are obtained through theory (solid lines), FEM ($\times$), and experiments (markers). The experiments were performed on the three beam specimens: s1-1, s1-2, and s1-3 (see Section~\ref{sec:exp_protocol}).}
  \label{fig:Fig6}
\end{figure}

We quantify the deformation of the beam by examining the deflections (normalized by the length of the beam) in the $x$ and $y$ directions, $\overline{\delta}_x=|\delta_x|/L$ (absolute value) and $\overline{\delta}_y=\delta_y/L$, respectively, and the rotation angle, $\theta_\mathrm{tip}$, at the tip. All of the results from the theory, FEM and experiments are plotted in Fig.~\ref{fig:Fig6} as a function of the level of magnetic loading, characterized by $\lambda_\mathrm{m}^\mathrm{C}$. The linear regime is observed in a narrow range $\lambda_\mathrm{m}^\mathrm{C}\lesssim2$. For larger values of $\lambda_\mathrm{m}^\mathrm{C}$, the deflections and rotation angle increase sublinearly with the applied field strength. This nonlinearity arises from the dependence of the magnetic body torque on the rotation angel of the beam ($\theta$); see Eq.~\eqref{magnetic_torque_BC}. As the rotation of the beam increases, its magnetization becomes increasingly aligned with the external field, thereby diminishing the magnitude of the magnetic body torque. Eventually, above a threshold $\lambda_\mathrm{m}^\mathrm{C}\approx20$, the deflections and rotation angle tend to a plateau. The tip angle reaches $90^\circ$, indicating the large rotations of the beam. This nonlinear behavior is accurately captured by our FEM and beam model (Fig.~\ref{fig:Fig6}), in excellent agreement with the experiments. 

\subsection{Hard-magnetic beam under a constant gradient magnetic field} 
\label{sec:gradient_field}

We now focus on the case of a constant gradient field, which has not been systematically investigated in previous work.
First, we specialize the beam model (Section~\ref{sec:model}) using Eqs.~\eqref{dBG_exp}, \eqref{BG_rp_exp}, and~\eqref{M_BG}, which describe the configurations of our experiments under a gradient field. In this case, the general forms of the magnetic body torque in Eq.~\eqref{m_torque_nondimensionless} and force in Eq.~\eqref{m_force_nondimensionless} reduce to 
\begin{equation}
\overline{\tau}_{z}^\mathrm{m}=-\frac{AL^2}{EI_\xi}\left[\nabla B_{rr} M \left({x}-\frac{1}{2}L\right) \cos{\theta}+\nabla B_{aa} M {y} \sin{\theta}\right]\,,
\label{magnetic_torque_BG}
\end{equation}
and
\begin{equation}
\overline{\hat{\mathbf{f}}}^\mathrm{m}=\frac{AL^3}{EI_\xi}\left(
-\nabla B_{rr} M\sin{\theta}, \nabla B_{aa} M\cos{\theta}\right)\,,
\label{magnetic_force_BG}
\end{equation}
respectively. The magnetic body force imposed on the beam arises from non-zero $\nabla\mathbf{B}$. Also, we take into account the gravitational force given in Eq.~\eqref{g_force_nondimensionless}, in the same way as in Section~\ref{sec:constant_field}.

We plug Eqs.~\eqref{magnetic_torque_BG} and~\eqref{magnetic_force_BG} of the magnetic loads and Eq.~\eqref{g_force_nondimensionless} of the gravitational force into the general equilibrium equation, Eq.~\eqref{Equi_Eq_nondimensionless}, obtaining
\begin{equation}
\begin{split}
0&=\theta_{,\bar{s}\bar{s}}\\
&+\frac{1}{2}\lambda_{\mathrm{m}}^{\nabla}\left[\cos{\theta}\left(2\int_{\bar{s}}^1{\cos{\theta}}\,\mathrm{d}\bar{s}'+\frac{x}{L}-\frac{1}{2}\right)-\sin{\theta}\left(\int_{\bar{s}}^{1}{\sin{\theta}}\,\mathrm{d}\bar{s}'+2\frac{y}{L}\right)\right]\\
&-(1-\bar{s})\overline{f}^\mathrm{g}_x\sin\theta\,.
\end{split}
\label{governing_equation_gradient}
\end{equation}
We note that we have made use of the relation $\nabla B_{rr}=-\frac{1}{2}\nabla B_{aa}$ for the Maxwell configuration. The boundary conditions of the cantilever beam are 
\begin{equation}
\theta(0)=0\,,\,\,\,\theta_{,\overline{s}}(1)=0\,,
\end{equation}
with which we can solve the equilibrium equation in Eq.~\eqref{governing_equation_gradient} to predict the deformed shape of the centerline of the beam. 

In Eq.~\eqref{governing_equation_gradient}, we recognize
\begin{equation}
\lambda_{\mathrm{m}}^{\nabla}=\frac{M \nabla B_{aa} AL^3}{EI_\xi}
\label{lambda_gradient}
\end{equation}
as the dimensionless magneto-elastic parameter for the case of a constant gradient field, which is analogous to $\lambda_{\mathrm{m}}^\mathrm{C}$ for the case of a uniform field. This parameter $\lambda_{\mathrm{m}}^{\nabla}$ quantifies the influence of magnetic loading on beam bending; \textit{i.e.}, the magneto-elastic coupling effect.

\begin{figure}[t!]
\centering
  \includegraphics[width=1\columnwidth]{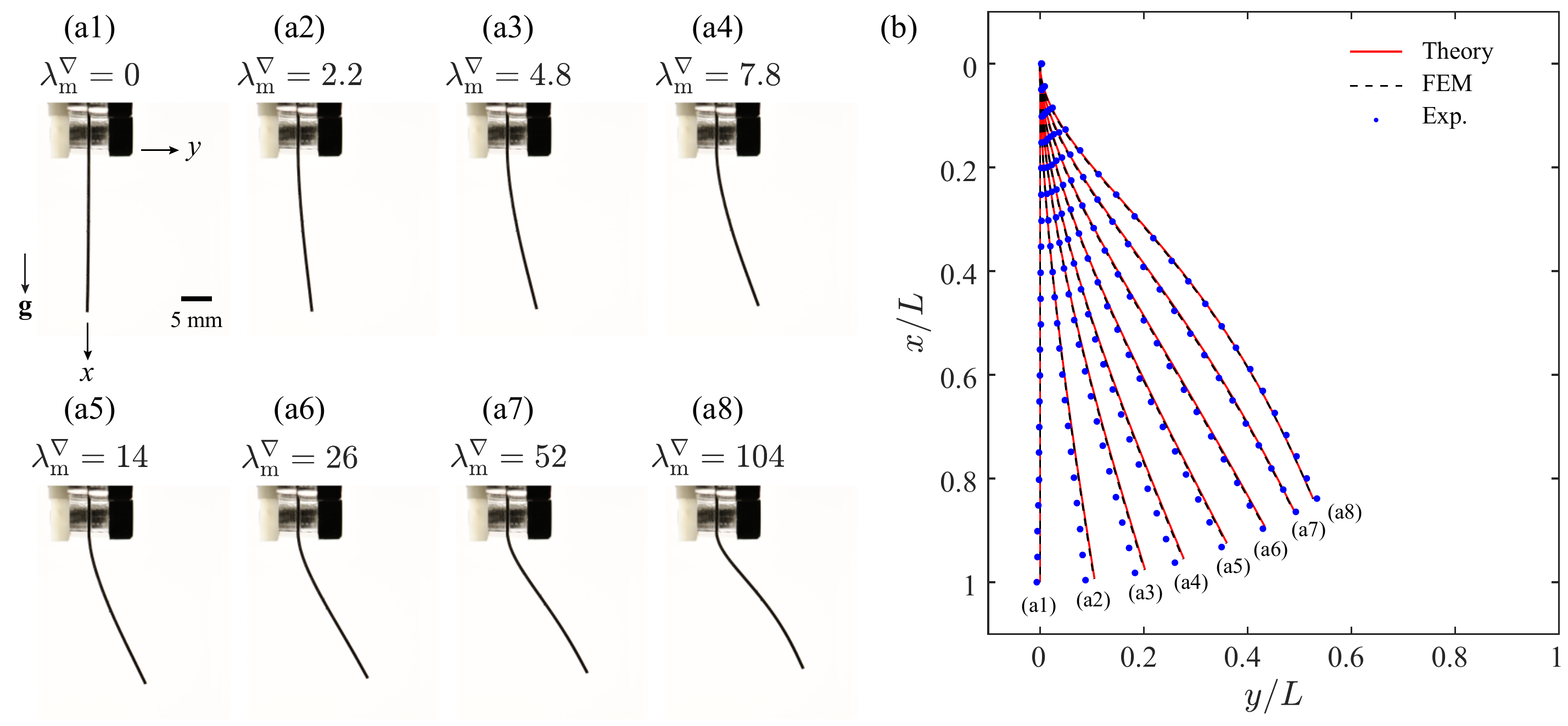}
  \caption{Configurations of a hard-magnetic beam subjected to a constant gradient magnetic field and gravity. The beam is magnetized in the $y$ direction in the initial configuration, parallel to the axial direction of the field generated by Maxwell coils (Section~\ref{sec:exp_protocol}). The gravitational field is in the $x$ direction. (a1)-(a8) Experimental photographs of the deformed beam at different levels of field gradients, quantified by the parameter $\lambda_{\mathrm{m}}^{\nabla}$; the respective values are indicated above each photograph. (b) Beam centerlines (data points) exacted from the photographs in (a), together with the corresponding predictions by the beam model (solid lines) and FEM (dashed lines).}
  \label{fig:Fig7}
\end{figure}

To analyze the deformation of the beam, in Fig.~\ref{fig:Fig7}(a1)-(a8), we present snapshots from the experiments at 8 values of $\lambda_{\mathrm{m}}^{\nabla}$. Fig.~\ref{fig:Fig7}(b) shows the corresponding beam centerlines extracted from the experimental photographs together with the predictions by the beam model and FEM. As the beam is actuated by the magnetic body force, it deflects toward the direction of increasing field strength ($y$). In Fig.~\ref{fig:Fig7}(a1)-(a5), for $\lambda_{\mathrm{m}}^{\nabla}\le14$, the shapes of the beam are similar to that observed in a uniform field in Fig.~\ref{fig:Fig5}(a1)-(a4); the beam undergoes large deflections and rotations. Counter-intuitively, in Fig.~\ref{fig:Fig7}(a6)-(a8), when further increasing the magnitude of field gradient above $\lambda_{\mathrm{m}}^{\nabla}>14$, the beam exhibits a distinct curved shape with a small increase in deflection at the tip.

\begin{figure}[t!]
\centering
  \includegraphics[width=1\columnwidth]{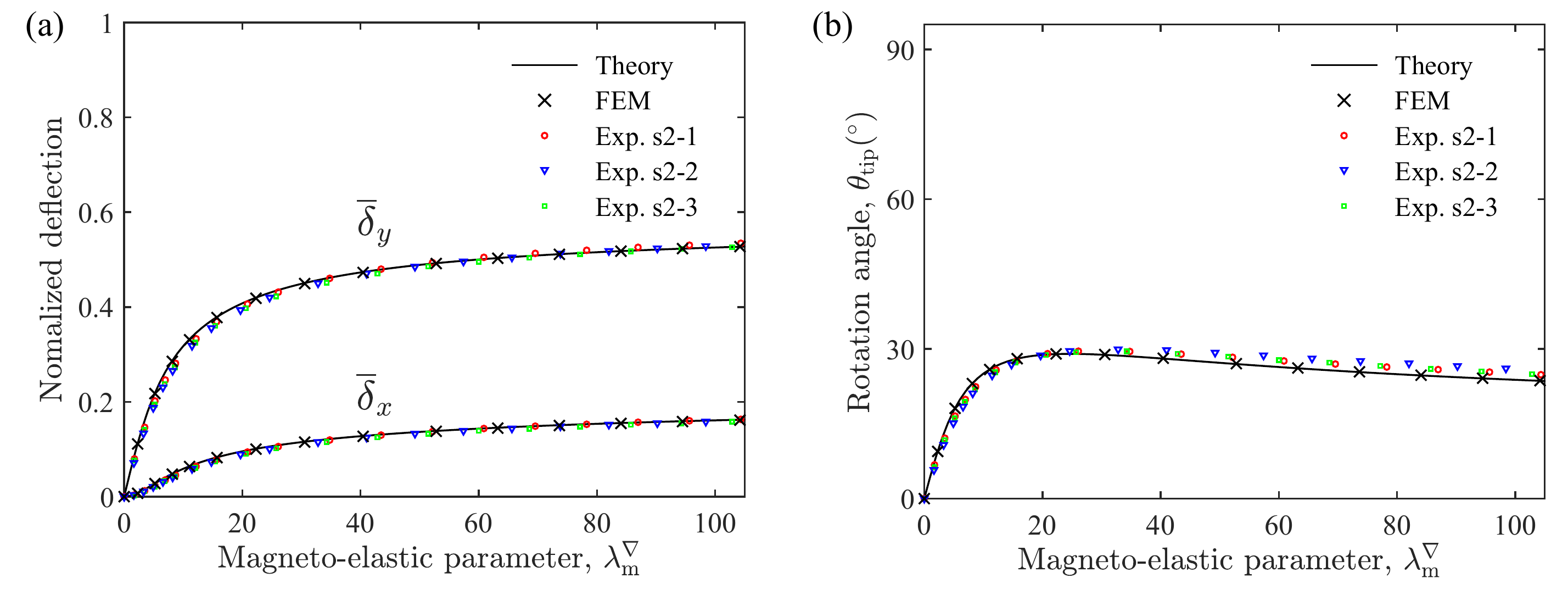}
  \caption{Beam deformation under a constant gradient magnetic field. (a) Normalized components of the deflection, $\overline{\delta}_x=|\delta_x|/L$ (absolute value) and $\overline{\delta}_y=\delta_y/L$, and (b) rotation angle, $\theta_\mathrm{tip}$, at the free end of the deformed beam versus the dimensionless magneto-elastic parameter $\lambda_{\mathrm{m}}^{\nabla}$. The results are obtained through theory (solid lines), FEM ($\times$), and experiments (markers). The experiments were performed on the three beam specimens: s2-1, s2-2, and s2-3 (Section~\ref{sec:exp_protocol}).} 
  \label{fig:Fig8}
\end{figure}

In Fig.~\ref{fig:Fig8}(a) and~(b), we plot the normalized deflections, $\overline{\delta}_x$ (absolute value) and $\overline{\delta}_y$, and rotation angle, $\theta_\mathrm{tip}$, at the tip of the beam, as functions of $\lambda_{\mathrm{m}}^{\nabla}$. For lower levels of applied field gradient $\nabla{B}_{aa}$ ($\lambda_{\mathrm{m}}^{\nabla}\lesssim25$), $\overline{\delta}_x$, $\overline{\delta}_y$ and $\theta_\mathrm{tip}$ increase linearly with $\lambda_{\mathrm{m}}^{\nabla}$, followed by a sublinear region. As the magnetic loading increases above $\lambda_{\mathrm{m}}^{\nabla}\gtrsim25$, the tip deflections reach a plateau. Meanwhile, beyond $\lambda_{\mathrm{m}}^{\nabla}\gtrsim25$, the rotation angle starts to decrease, indicating that the tip of the beam rotates in the opposite direction (clockwise); see Fig.~\ref{fig:Fig7}(a6)-(a8). The plateau of deflection and the reverse of the direction of rotation are due to the magnetic body torque, which increases with the rotation angle but acts against beam bending.

Compared to the beam under a uniform field (Fig.~\ref{fig:Fig6}), with increasing magnetic loading, we observed a much lower plateau of deflection and a non-monotonic rotation angle at the tip under a constant gradient field (Fig.~\ref{fig:Fig8}).
We highlight that, in Figs.~\ref{fig:Fig7}(b) and~\ref{fig:Fig8}, the juxtapositions of the experimental, theoretical, and computational results validate that the beam model (Section~\ref{sec:model}) and FEM (Section~\ref{sec:theory_FEM}) proposed for the case of a constant gradient field can quantitatively predict the deformation of hard-magnetic beams.

\section{Comparative studies for actuation under uniform versus gradient magnetic fields}
\label{sec:comparison}

The framework established in Sections~\ref{sec:model} and~\ref{sec:theory_FEM}, validated experimentally in Section~\ref{sec:results}, now enables comparative studies on hard-magnetic beams in different field configurations. In this last section, we focus on the developed reduced-order theory and 3D FEM to investigate two case studies toward gaining additional insight into the response of hard-magnetic beams. In case study (i), we compare the deflections of a cantilever beam actuated by either a uniform or a constant gradient field. In case study (ii), under a magnetic field, we examine the stiffness of a cantilever beam in response to a mechanical point load imposed at the tip. These comparative studies serve to illustrate the versatility of the developed framework. The results provide quantitative comparisons in specific cases between the efficacy of uniform and gradient fields.

\subsection{Beam deflection in different field configurations}
\label{sec:case_study_i}

\begin{figure}[t!]
\centering
 \includegraphics[width=\columnwidth]{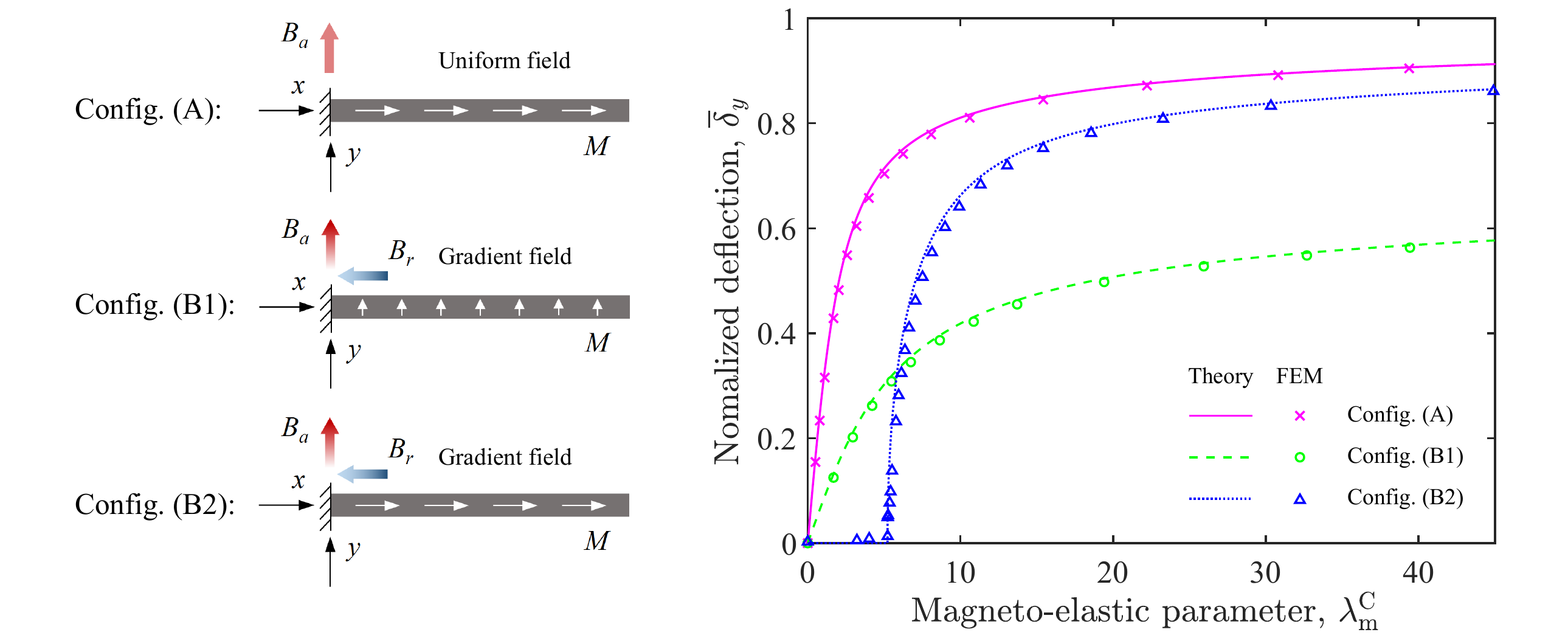}
 \caption{Normalized tip deflection of a cantilever beam, $\overline{\delta}_y$, versus the governing parameter, $\lambda_\mathrm{m}^\mathrm{C}$, in Configurations~(A), (B1), and~(B2). The uniform and constant gradient fields are generated by the same pair of coils. The ratio $L/R_\mathrm{c}=0.4$ is fixed in Configurations~(B1) and~(B2). The results are obtained from both the beam model (lines) and FEM (data points).}
 \label{fig:Fig9}
\end{figure}

We start with case study (i), in which beams made of the same hard-MRE and with the same dimensions are tested in three configurations, as depicted in Fig.~\ref{fig:Fig9}(a). In Configuration~(A), the beam is magnetized in the length-wise direction and subjected to a uniform field perpendicular to the magnetization. Magnetic torques are then generated to deflect the beam. This configuration is identical to that used in Section~\ref{sec:constant_field} for the experimental validation. In Configurations~(B1) and~(B2), we apply a constant gradient field on the beam, which is magnetized in either the thickness or the length direction, respectively. The axis of the applied field is perpendicular to the beam centerline, and the center of the field is aligned with the fixed end of the beam; \textit{i.e.}, $\mathbf{x}_\mathrm{p}=(0,0,0)$. In these two configurations, (B1) and~(B2), under a gradient field, the beam deflects due to a combination of magnetic torques and forces. 

To further simplify this comparative study, we neglect the effect of gravity. As such, according to Eq.~\eqref{Equi_Eq_nondimensionless}, the equilibrium equations of the beam in Configurations~(A), (B1), and~(B2) are, respectively,
\begin{equation}
\theta_{,\bar{s}\bar{s}}+\lambda_\mathrm{m}^\mathrm{C}\cos{\theta}=0\,,
\label{beam_config_a}
\end{equation}
\begin{equation}
\theta_{,\bar{s}\bar{s}}+\frac{1}{2}\lambda_{\mathrm{m}}^{\nabla}\left[\cos{\theta}\left(2\int_{\bar{s}}^1{\cos{\theta}}\,\mathrm{d}\bar{s}'+\frac{x}{L}\right)-\sin{\theta}\left(\int_{\bar{s}}^{1}{\sin{\theta}}\,\mathrm{d}\bar{s}'+2\frac{y}{L}\right)\right]=0\,,
\label{beam_config_b1}
\end{equation}
and
\begin{equation}
\theta_{,\bar{s}\bar{s}}+\frac{1}{2}\lambda_{\mathrm{m}}^{\nabla}\left(2\cos{\theta}\int_0^1{\sin{\theta}}\,\mathrm{d}\bar{s}+\sin{\theta}\int_{0}^{1}{\cos{\theta}}\,\mathrm{d}\bar{s}\right)=0\,.
\label{beam_config_b2}
\end{equation}
The deformation is governed by a single parameter: either $\lambda_\mathrm{m}^\mathrm{C}$, Eq.\eqref{lambda_uniform} for Configuration~(A), or $\lambda_{\mathrm{m}}^{\nabla}$, Eq.\eqref{lambda_gradient} for Configurations~(B1) and~(B2).

We compare the deflections of the beam in the three configurations by assuming that the uniform and constant gradient fields are generated by the same pair of coils (same values of current $I_\mathrm{c}$, radius $R_\mathrm{c}$, number of turns $n_\mathrm{c}$), set either in the Helmholtz or Maxwell configuration. Derived from the Biot-Savart law~\citep{grant_electromagnetism_1990}, when coils are in the Helmholtz configuration, the flux density of the generated uniform field in the axial direction is 
\begin{equation}
B_{a}=\left(\frac{4}{5}\right)^{\frac{3}{2}}\frac{\mu_0n_\mathrm{c}I_\mathrm{c}}{R_\mathrm{c}}\,.
\label{B_H}
\end{equation}
Whereas, when coils are in the Maxwell configuration, the gradient of $B_{a}$ along the axis is
\begin{equation}
\nabla{B}_{aa}=\frac{3\sqrt{3}}{2}\left(\frac{4}{7}\right)^{\frac{5}{2}}\frac{\mu_0n_\mathrm{c}I_\mathrm{c}}{R_\mathrm{c}^2}\,,
\label{dB_M}
\end{equation}
Then, from Eqs.~\eqref{lambda_uniform} and~\eqref{lambda_gradient}, the governing parameters $\lambda_\mathrm{m}^\mathrm{C}$ and $\lambda_{\mathrm{m}}^{\nabla}$ are related as 
\begin{equation}
\lambda_{\mathrm{m}}^{\nabla}\approx0.90\frac{L}{R_\mathrm{c}}\lambda_\mathrm{m}^\mathrm{C}\,,
\label{lambda_uniform_gradient}
\end{equation}
where ${L}/{R_\mathrm{c}}$ is the ratio between length of the beam and radius of the coils.

Using Eq.~\eqref{lambda_uniform_gradient}, we are now able to plot the tip deflections of the beam, $\overline{\delta}_y$, in Configurations~(A), (B1), and~(B2) versus the same parameter, $\lambda_\mathrm{m}^\mathrm{C}$, as shown in Fig.~\ref{fig:Fig9}. The curves for Configurations~(B1) and~(B2) depend also on ${L}/{R_\mathrm{c}}$, which is fixed at $L/R_\mathrm{c}=0.4$. For completeness, the influence of $L/R_\mathrm{c}$ on the comparison is discussed in \ref{appendix_A}. We can see from the data in Fig.~\ref{fig:Fig9} that, for a given value of $\lambda_\mathrm{m}^\mathrm{C}$, the tip deflection of the beam in Configuration~(A) driven by magnetic torques is always larger than that in Configuration~(B1) driven by magnetic forces. In Section~\ref{sec:gradient_field}, we have already mentioned a rationale underlying this observation: in Configuration~(B1), magnetic torques counteract the rotation of the beam, resulting in a much lower plateau of $\overline{\delta}_y$. A distinct behavior of the beam is observed in Configuration~(B2): the beam remains in the straight configuration ($\theta=0$) as the field gradient increases until $\lambda_\mathrm{m}^\mathrm{C}\approx5.2$, at which point buckling occurs (bending becomes energetically more favorable, compared to the straight configuration). Since both torques and forces facilitate bending deformation, the tip deflection increases with $\lambda_\mathrm{m}^\mathrm{C}$ at a higher rate for Configuration~(B2) than for Configurations~(A) and~(B1). 

These observations presented above indicate that, when a field is applied perpendicularly to a beam that is magnetized along the centerline, a uniform field is a preferable choice for actuation in terms of the extent of deflection and the required energy input from electromagnetic coils, while a constant gradient field could induce buckling of the beam. We note that, in the FEM simulation for Configuration~(B2), a small perturbation (constant) magnetic field perpendicular to the beam was imposed to trigger the transition at the critical point, from the straight to the curved configuration.

\subsection{Change in bending stiffness of the beam by applying a magnetic field}
\label{sec:case_study_ii}

\begin{figure}[t!]
\centering
 \includegraphics[width=\columnwidth]{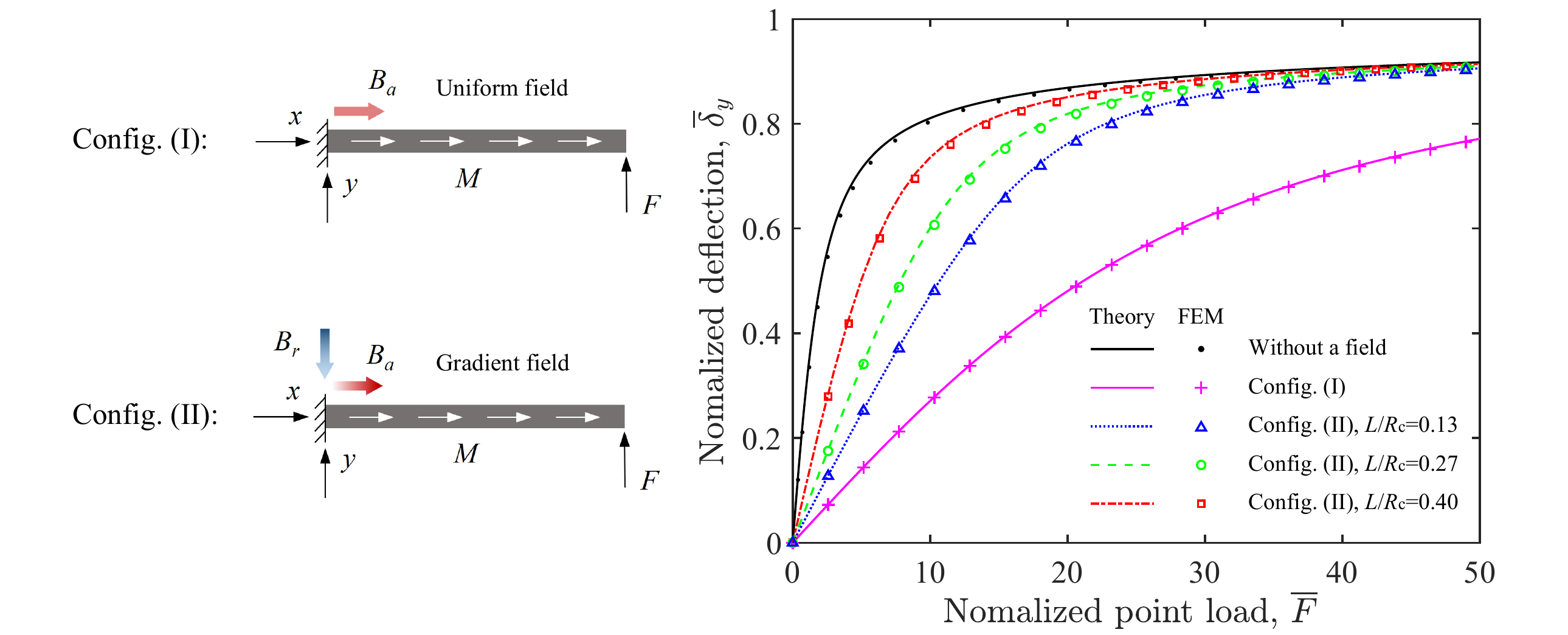}
 \caption{Normalized tip deflection of a cantilever beam, $\overline{\delta}_y$, versus the normalized point load, $\overline{F}=FL^2/EI$, applied at the tip. The beam is loaded by point force ${F}$ at its tip under a uniform field in Configuration~(I) or a constant gradient field in Configuration~(II). The responses are compared to the case without an applied field. The fields are generated by the same pair of coils and the field strength is fixed at $\lambda_\mathrm{m}^\mathrm{C}=28.6$. The ratio $L/R_\mathrm{c}=\{0.13,0.27,0.40\}$ is varied. The results are obtained from both the beam model (lines) and FEM (data points).}
 \label{fig:Fig10}
\end{figure}

In case study (ii), we consider a cantilever beam subjected to a point load, $F$, at the tip, and examine how the stiffness of the beam is affected by either a uniform or a gradient field. Configurations~(I) and~(II) of the beam and the applied fields are schematized in Fig.~\ref{fig:Fig10}. We align the beam centerline with the axis of the coils, along which the beam is magnetized. Also, to avoid buckling~\citep{dehrouyeh-semnani_bifurcation_2021}, the fixed end of the beam is positioned at the center of the field, \textit{i.e.}, $\mathbf{x}_\mathrm{p}=(0,0,0)$, such that the magnetization is parallel to the field in the axial direction along the entire beam. We specialize Eq.~\eqref{Equi_Eq_nondimensionless} according to Configurations~(I) and~(II), obtaining the equilibrium equations of the beam

\begin{equation}
\theta_{,\bar{s}\bar{s}}-\lambda_\mathrm{m}^\mathrm{C}\sin{\theta}+\frac{F L^2}{EI}\cos{\theta}=0\,,
\end{equation}
and
\begin{equation}
\theta_{,\bar{s}\bar{s}}-\frac{1}{2}\lambda_{\mathrm{m}}^{\nabla}\left(\cos{\theta}\int_0^1{\sin{\theta}}\,\mathrm{d}\bar{s}+2\sin{\theta}\int_{0}^{1}{\cos{\theta}}\,\mathrm{d}\bar{s}\right)+\frac{F L^2}{EI}\cos{\theta}=0\,,
\end{equation}
respectively.
 
In the initial configuration, we first apply a magnetic field at a given level of field strength. In the uniform field, there is no magnetic torques imposed on the beam, since the magnetization and the field are parallel. In the constant gradient field, the net torque acting on the cross-section of the beam is zero. Magnetic forces are generated slightly stretching the beam, which is taken in account in the FEM but ignored in the reduced-order beam model due to the assumption of inextensibility. Hence, under both fields, the beam is at rest in the straight configuration.

Once we apply a point load at the tip to deflect the beam, magnetic toques and/or forces come into play. In Fig.~\ref{fig:Fig10}, we plot the normalized tip deflection of the beam, $\overline{\delta}_y$, versus the normalized point load, $\overline{F}=FL^2/EI$, for the following three situations: (I) a uniform field, (II) a constant gradient field, and (III) without an applied field. Again, the fields are generated by the same coils, and $\lambda_\mathrm{m}^\mathrm{C}=28.6$ is fixed throughout the comparison. The slopes of the curves correspond to the compliance of the beam in response to the point load. We find that, in either Configuration~(I) or~(II), the beam becomes less compliant, with respect to the case in the absence of a field. These findings demonstrate that, by applying a magnetic field, it is possible to modify the effective bending stiffness of the beam against mechanical loads. Comparing the change in stiffness under the two fields, the uniform field is more performant for all values of $L/R_\mathrm{c}=\{0.13,0.27,0.40\}$ explored. From the curves in Fig.~\ref{fig:Fig10}, it is also clear that the stiffness of the beam in the constant gradient field increases with $L/R_\mathrm{c}$ for a given value of $\lambda_\mathrm{m}^\mathrm{C}$. However, further increasing $L/R_\mathrm{c}$ by using a longer beam or coils with smaller radii, the beam would reach into the region of non-uniform gradient between the coils. To consider this non-linearity of the field, additional work on the beam model and FEM would be needed, which we hope future research will address.

\subsection{Discussion on the comparison between uniform versus gradient fields}
\label{sec:case_study_discussion}

Through the two case studies in Sections~\ref{sec:case_study_i} and~\ref{sec:case_study_ii}, we have demonstrated the ability to systematically study the deformation of hard-magnetic beams under uniform and gradient fields, using the proposed theory and FEM frameworks. The uniform field outperforms the gradient field in deflecting a beam and in changing the stiffness of a beam in response to mechanical loads. However, the inhomogeneity of a gradient field could enrich the beam deformation mode (\textit{e.g.}, induce buckling). Also, a gradient field provides additional degrees of freedom (gradients) for the design and control of sophisticated shape transformations or locomotion of a hard-magnetic body~\citep{Lum_PNAS2016,Hu_Nature2018,diller_six-degree--freedom_2016,kummer_octomag_2010}. With the understanding of the effect of field gradients on deformable beam-like structures, we anticipate that, in the future, the generated magnetic force can be better leveraged in pulling, transport, navigation, and manipulation for applications in soft robotics and biomedicine~\citep{li_magnetic_2016,yesin_modeling_2006,Kim_Nature2018,wu_multifunctional_2020,dong_controlling_2020}. 

\section{Conclusion}
\label{sec:conclusion}

We have developed a comprehensive framework to predict the response of hard-magnetic beams under magnetic actuation, with a focus on non-uniform external fields with a constant gradient, of which the uniform field case is a simple limit. A reduced-order model was derived rigorously, starting from the 3D Helmholtz free energy of the beam, through dimensional reduction. We also extended the existing 3D theory for hard-magnetic solids to take external field gradients into account. Both the reduced-order beam theory and the 3D FEM simulations were validated through precision experiments under either uniform or gradient fields.

In a specific test case, we studied the deformation of a cantilever beam under magnetic loading. In a \textit{uniform field}, the beam is deflected by magnetic torques. The deflection and rotation are nonlinear with the applied field, which are governed by a single magneto-elastic parameter that characterizes the coupling between bending deformation and magnetic actuation. In a \textit{constant gradient field}, the beam deflection is driven by magnetic forces but resisted by magnetic torques. As a result, the plateau of deflection is much lower than that under a uniform field, and the direction of rotation at the tip is reversed above a threshold level of the field gradient. A second magneto-elastic parameter is identified for this configuration. 

We carried out a set of comparative case studies for a cantilever beam in different configurations. For a fixed input power from the coils, a uniform field yields a larger deflection of the beam compared to a constant gradient field. The beam magnetized in the length direction undergoes buckling from the straight to a curved shape, when subjected to a constant gradient field (field axis perpendicular to the beam centerline). In a second case study, we demonstrated that applying a field parallel to the magnetization vector could stiffen the beam in response to a point load imposed at the tip. In terms of the magnitude of the change in stiffness, a uniform field outperforms a constant gradient field. These comparative studies illustrate the versatility of the proposed framework.

Our findings presented in this study provide valuable fundamental insight into the magneto-elastic coupling effect in thin beams. The buckling behavior of hard-magnetic beams and their ability for stiffness tuning could be leveraged to devise functionality for applications. Methodologically, the theoretical and computational tools that we developed could be employed for the predictive and rational design of magnetically-active systems comprising beam elements. In particular, the dimensional reduction procedure demonstrated here for thin beams can be followed to develop reduced-order models for other slender structural elements, such as rods, plates, and shells~\citep{yan_magneto-active_2021}. We note that, ~\cite{sano_rods_2021} have recently proposed a Kirchhoff-like theory for hard magnetic rods under geometrically nonlinear deformation in three dimensions. More generally, the present 3D FEM procedure  allows for simulations on hard-magnetic structures with arbitrary geometries. 

It is important to note that the magnetic dipole-dipole interactions between the magnetized particles embedded in our MRE were neglected throughout the present study. This simplification was justified by the excellent agreement between the predictions and experimental measurements. However, for even softer matrix materials than those we considered, the dipole-dipole interactions may play an important role in the mechanical properties and response of hard-MREs, as has been recently shown in the literature~\citep{vaganov_effect_2018,vaganov_training_2020,schumann_reversible_2021,zhang_micromechanics_2020,garcia-gonzalez_microstructural_2021}. The influence of non-local magnetic interactions on the deflection of thin beams is another interesting topic for future research.

\appendix

\section{Influence of ${L}/{R_\mathrm{c}}$ on the comparison between the beam deflections in a uniform and a gradient field}
\label{appendix_A}

In the comparative case study presented in Section~\ref{sec:case_study_i}, we compared the deflections of a cantilever beam in Configurations~(A), (B1), and~(B2) by plotting $\overline{\delta}_y$ versus the parameter $\lambda_\mathrm{m}^\mathrm{C}$, defined in Eq.~\eqref{lambda_uniform}, in Fig.~\ref{fig:Fig9}. Since the beam deflection under a gradient field is governed by $\lambda_{\mathrm{m}}^{\nabla}$, defined in Eq.~\eqref{lambda_gradient}, and related with $\lambda_\mathrm{m}^\mathrm{C}$ through Eq.~\eqref{lambda_uniform_gradient}, the curves for Configurations~(B1) and~(B2) depend on ${L}/{R_\mathrm{c}}$, the ratio between length of the beam and radius of the coils. In Fig.~\ref{fig:Fig9}, we fixed ${L}/{R_\mathrm{c}}=0.4$.

\begin{figure}[h!]
\centering
 \includegraphics[width=\columnwidth]{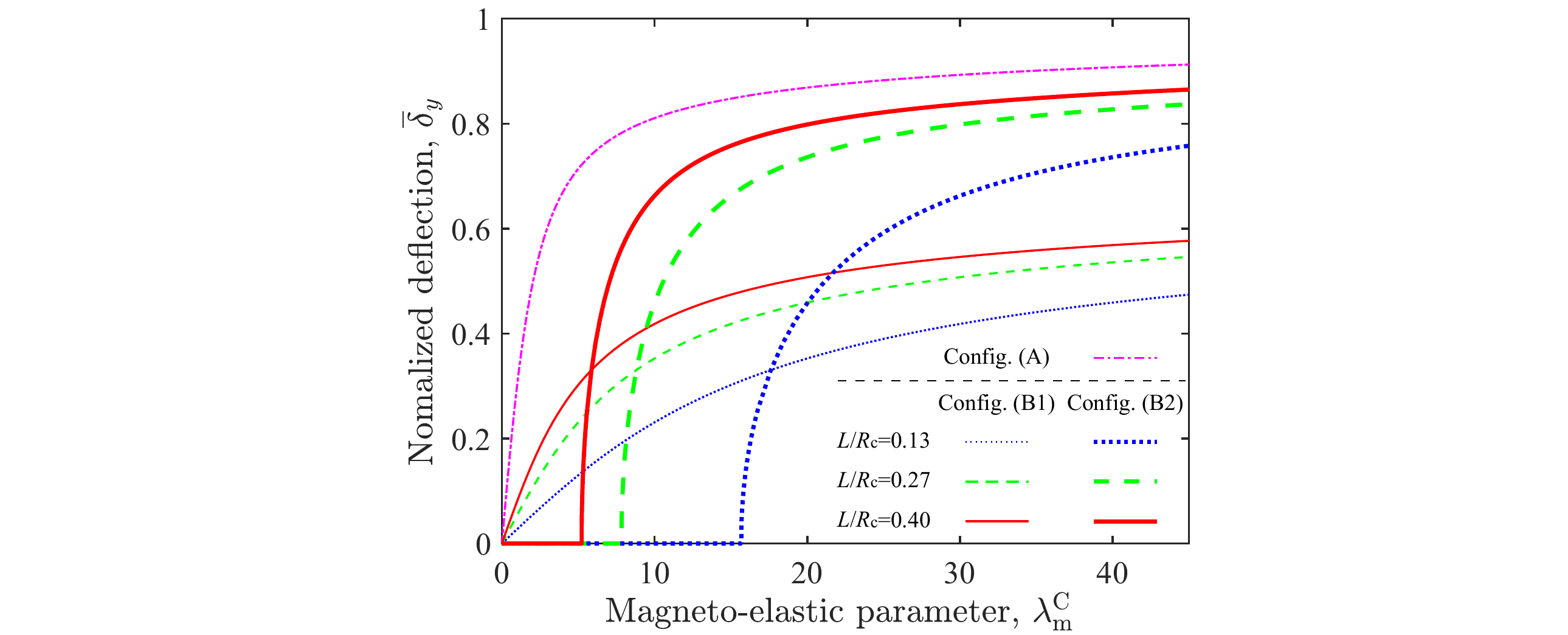}
 \caption{Normalized tip deflection of a cantilever beam, $\overline{\delta}_y$, versus the governing parameter, $\lambda_\mathrm{m}^\mathrm{C}$, in Configurations~(A), (B1), and~(B2) schematized in Fig.~\ref{fig:Fig9}. The ratio $L/R_\mathrm{c}=\{0.13,0.27,0.40\}$ is varied to evaluate its influence on the beam deflections in Configurations~(B1) and~(B2). The results are predicted by the reduced-order beam model developed in Section~\ref{sec:model}.}
 \label{fig:FigA11}
\end{figure}

To more thoroughly evaluate the influence of $L/R_\mathrm{c}$, in Fig.~\ref{fig:FigA11}, we provide the normalized tip deflection, $\overline{\delta}_y$, versus $\lambda_\mathrm{m}^\mathrm{C}$, while varying $L/R_\mathrm{c}=\{0.13,0.27,0.40\}$. This range of $L/R_\mathrm{c}$ ensures that the beam remains within the region of constant gradient between the two coils (see Fig.~\ref{fig:Fig4}). We find that, for a given value of $\lambda_\mathrm{m}^\mathrm{C}$, the tip deflections in Configurations~(B1) and~(B2), under a gradient field, increase with $L/R_\mathrm{c}$. However, in the range of $L/R_\mathrm{c}$ explored, $\overline{\delta}_y$ under a uniform field is always larger than a gradient field. 

\bigskip
\noindent \textbf{Acknowledgments:} A.A. is grateful to the support from the Federal Commission for Scholarships for Foreign Students (FCS) through a Swiss Government Excellence Scholarship (Grant No. 2019.0619). We thank Matteo Pezzulla for fruitful discussions, and Lilian Cruveiller for help with preliminary explorative experiments.

\bibliography{references}

\end{document}